\documentclass[aps,prd,floats,floatfix,nofootinbib,preprintnumbers]{revtex4-1}

\pdfoutput=1

\usepackage{graphicx,amsmath}
\usepackage{amssymb}
\usepackage{amsfonts}
\usepackage{hyperref}
\usepackage[bottom]{footmisc}
\usepackage{subfigure}

\begin{document}

\preprint{FERMILAB-PUB-11-017-T}

\title{Top-Higgs and Top-pion phenomenology in the Top Triangle Moose model}

\author{R.\ Sekhar Chivukula}
\email {sekhar@msu.edu}
\affiliation {Department of Physics and Astronomy,
Michigan State University,
East Lansing, MI 48824, USA}

\author{Baradhwaj Coleppa}
\email {barath@physics.carleton.ca }
\affiliation {Ottawa-Carleton Institute for Physics,
Carleton University,
Ottawa, Ontario K1S 5B6, Canada}

\author{Heather E.\ Logan}
\email {logan@physics.carleton.ca }
\affiliation {Ottawa-Carleton Institute for Physics,
Carleton University,
Ottawa, Ontario K1S 5B6, Canada}

\author{Adam Martin}
\email {aomartin@fnal.gov }
\affiliation {Theoretical Physics Department,
Fermilab, 
Batavia, IL 60510, USA}

\author{Elizabeth H.\ Simmons}
\email {esimmons@pa.msu.edu}
\affiliation {Department of Physics and Astronomy,
Michigan State University,
East Lansing, MI 48824, USA}

\date{\today}

\begin{abstract}
We discuss the deconstructed version of a topcolor-assisted
technicolor model wherein the mechanism of top quark mass generation
is separated from the rest of electroweak symmetry breaking. The
minimal deconstructed version of this scenario is a ``triangle moose''
model, where the top quark gets its mass from coupling to a top-Higgs
field, while the gauge boson masses are generated from a Higgsless
sector. The spectrum of the model includes scalar (top-Higgs) and
pseudoscalar (top-pion) states. In this paper, we study the properties
of these particles, discuss their production mechanisms and decay modes, and suggest how best to search for them at the LHC.
\end{abstract}

\maketitle

\section{Introduction}

Higgsless models, as the name implies, break the electroweak symmetry without invoking a fundamental scalar particle. Inspired by the idea that one could maintain perturbative unitarity in extra-dimensional models through heavy vector resonance exchanges in lieu of a Higgs \cite{SekharChivukula:2001hz,Chivukula:2002ej,Chivukula:2003kq}, Higgsless models were intially introduced in an extra-dimensional context as $SU(2)\times SU(2)\times U(1)$ gauge theories living in a slice of $AdS_5$, with symmetry breaking codified in the boundary condition of the gauge fields \cite{Csaki:2009bb,Cacciapaglia:2006gp,Cacciapaglia:2004zv,Csaki:2004sz,Cacciapaglia:2004rb,Cacciapaglia:2004jz,Csaki:2003zu}. It emerged that the low energy dynamics of these extra-dimensional models can be understood in terms of a collection of 4-D theories, using the principle of ``deconstruction'' \cite{Hill:2000mu,ArkaniHamed:2001ca}. Essentially, this involves latticizing the extra dimension, associating a 4-D gauge group with each lattice point and  connecting them to one another by means of nonlinear sigma models. The five dimensional gauge field is now spread in this theory as four dimensional gauge fields residing at each lattice point, and the fifth scalar component residing as the eaten pion in the sigma fields. The picture that emerges is called a ``Moose'' diagram~\cite{Georgi:1985hf}. The $AdS/CFT$ correspondence suggests that these models can be understood to be dual to strongly coupled technicolor models. The key features of these models \cite{Casalbuoni:1985kq,Casalbuoni:1995qt,Chivukula:2005ji,Chivukula:2005xm,Chivukula:2005bn,Kurachi:2004rj,Chivukula:2004af,Chivukula:2004pk,Lane:2009ct} are the following:
\begin{itemize}
\item Spin-1 resonances created by the techni-dynamics are described
  as massive gauge bosons, following the Hidden-Local-Symmetry setup
  originally used for
  QCD~\cite{Bando:1987br,Bando:1985rf,Bando:1984pw,Bando:1984ej,Bando:1987ym}.
  The mass of the resonances is roughly $\tilde g F$, where $F$ is
  around the weak scale and $\tilde g$ is a large
  coupling. Interactions of the resonances with other gauge bosons and
  fermions are calculated as a series in $1/\tilde g \ll 1$.
\item Standard model (SM) fermions reside primarily on the exterior sites -- the
  sites approximately corresponding to $SU(2)_w$ and $U(1)_Y$ gauge groups. Fermions
  become massive through mixing with massive, vector-like fermions
  located on the interior, `hidden' sites.
\item Precision electroweak parameters (S,T,U), perennially a thorn in
  the side of dynamical electroweak breaking
  models~\cite{Peskin:1991sw}, are accommodated by delicately
  spreading the SM fermions between sites. By adjusting the fermion
  distribution across sites to match the gauge boson distribution, S,
  T, U can all be reduced to acceptable levels. This is identical to
  the solution used in extra-dimensional Higgsless
  models,
  where the spreading of a fermion among sites becomes a continuous
  distribution, or profile, in the extra dimension~\cite{Cacciapaglia:2006gp}.  This adjustment is
  called ``ideal delocalization'' \cite{Chivukula:2005xm}.
\end{itemize}

The most economical deconstructed Higgsless model constructed along these lines (a ``three site'' model) was presented in \cite{Chivukula:2006cg}, and had, in addition to the SM spectrum, one heavy partner for each fermion and the $W$ and $Z$ bosons. Though providing an excellent ground for studying the low-energy properties of Higgsless models, the mass of the heavy Dirac partners of the SM fermions in this model was constrained to lie at $\sim$ 2 TeV, because of the tension between obtaining the correct value for the top-quark mass and keeping $\Delta \rho$ within experimental bounds. To alleviate this constraint, an extension of the three site model was constructed \cite{Chivukula:2009ck}, whose goal was to separate the top-quark mass generation from the rest of electroweak symmetry breaking (EWSB), thus relaxing the aforementioned constraint.

 This idea of treating the top-quark mass as arising from a separate dynamics is not new - in fact, Topcolor-assisted technicolor models \cite{Hill:1991at,Hill:1994hp,Lane and Eichten - TC2,Popovic:1998vb, Hill and Simmons, Braam:2007pm} employ precisely this idea. Topcolor-assisted technicolor is a
scenario of dynamical electroweak symmetry breaking in which the
strong dynamics is partitioned into two different sectors. One sector,
the technicolor sector, is responsible for the bulk of electroweak
symmetry breaking and is therefore characterized by a scale $F \sim
v$, where $v=$ 246 GeV is the EWSB scale. Consequently, technicolor dynamics is responsible for the majority
of gauge boson masses and, more indirectly, light fermion masses. 
The second strong sector, the topcolor sector, only communicates
directly with the top quark. Its sole purpose is to generate a large
mass for the top quark. In generating a top quark mass, this second sector
also breaks the electroweak symmetry. If the characteristic scale of the
topcolor sector is low, $f \ll F$, it plays only a minor role in
electroweak breaking, but can still generate a sufficiently large top quark mass
given a strong enough top-topcolor coupling. At low-energies, the top-color dynamics is summarized by the existence of a new dynamical top-Higgs which couples preferentially to the top-quark. The introduction of the top-Higgs field serves to alleviate the tension between obtaining
the correct top quark mass and keeping $\Delta \rho$ small that exists
in Higgsless models by separating the top quark mass generation from
the rest of electroweak symmetry breaking. An important
consequence of this separation is that the model permits heavy Dirac partners for
the fermions that are potentially light enough to be seen at the LHC\@.
Thus, the combination of two symmetry-breaking mechanisms can
achieve both dynamical electroweak breaking and a realistic top quark
mass. 

Because electroweak symmetry is effectively broken twice in this
scenario, there are two sets of Goldstone bosons in the theory. One
triplet of these Goldstones is eaten to become the longitudinal modes
of the $W^{\pm}/Z^0$, while the other triplet remains in the
spectrum. This remaining triplet, typically called the top-pions, and
a singlet partner, the top-Higgs, are the focus of this paper.

The top-pions and top-Higgs couple preferentially to the third
generation of quarks, which makes them interesting for a number of
reasons. First, the interactions of the top quark are the least
constrained of all fermions, so new dynamics coupling preferentially to the top quark is not phenomenologically excluded. Second, the gluon fusion mechanism involves a top quark loop
and is an efficient method for singly producing top-Higgses and
neutral top-pions at the LHC. In fact, the strong top-quark--topcolor interaction, manifest in a top Yukawa of order $\sim$ few, significantly
enhances the coupling of top-Higgs and top-pions relative to a SM Higgs of equal mass . Such a large cross section leads to exciting LHC signals
which may be discoverable in the initial low-energy, low-luminosity
run.


Our goal in this paper is to lay the foundation for phenomenological
studies of the top-pions and top-Higgs at the LHC.
We begin in Sec.~\ref{sec:model} by setting out the relevant details of
the Top Triangle Moose model.  In Sec.~\ref{sec:couplings} we identify
the physical top-pion states and summarize their couplings to other
particles.  Section~\ref{sec:pheno} contains the bulk of our
phenomenological results.  After identifying the existing experimental
constraints on the top-pions and top-Higgs, we study their decay
branching ratios, direct production cross sections in $pp$ collisions,
and production in decays of the heavy vector-like top quark
partners or the heavy gauge bosons in the model.  We identify cases 
where the LHC has the clear ability to discover the new states and others 
where good potential for discovery exists and further detailed study is 
warranted. We summarize our findings and discuss their implications in 
Sec.~\ref{sec:conclusions}. 

\section{The model}
\label{sec:model}

\begin{figure}[h!]
\begin{center}
\includegraphics[angle=0,width=2.4in]{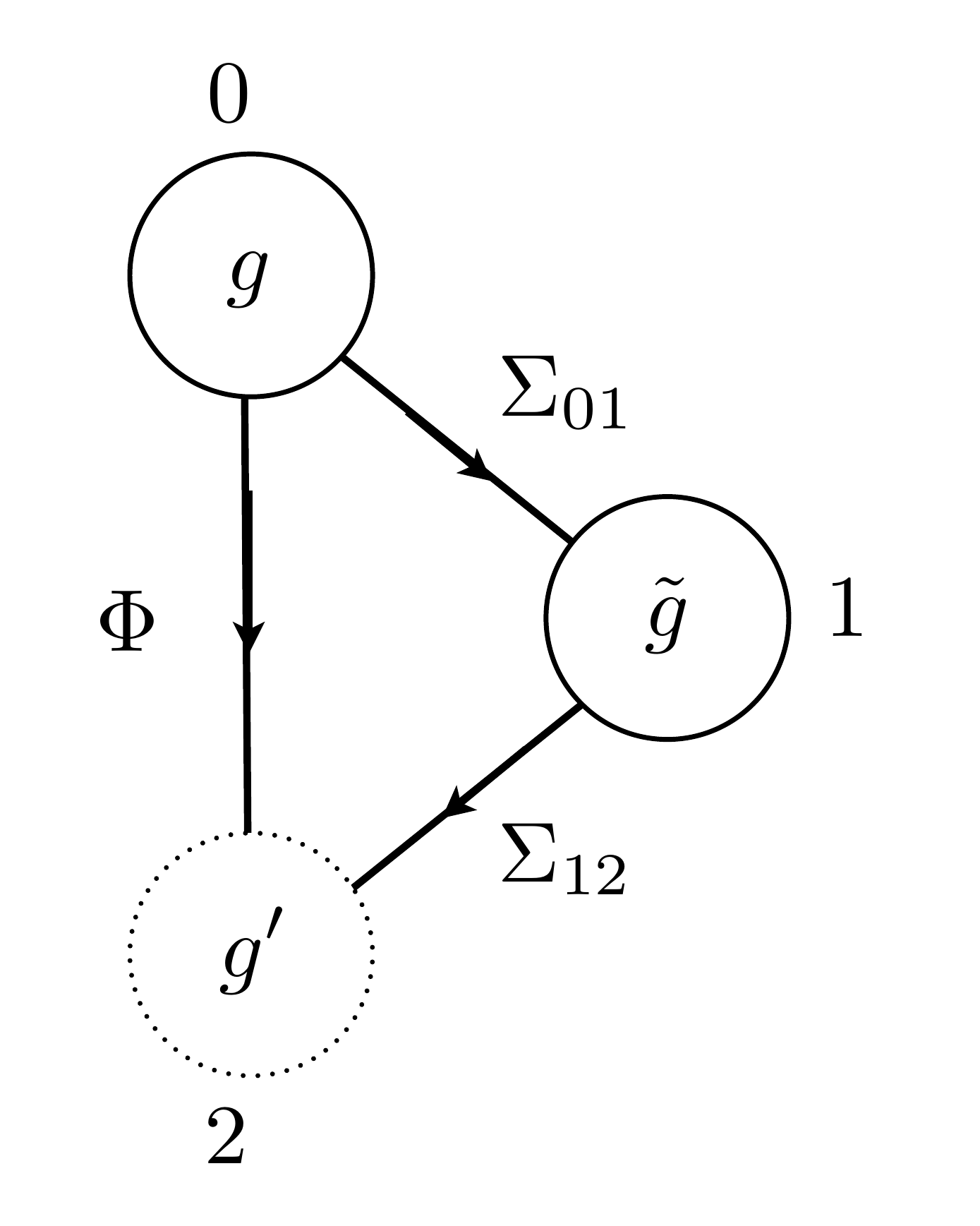}
\caption{The gauge structure of the model in Moose notation. $g$ and
  $g^{\prime}$ are approximately the SM $SU(2)$ and hypercharge gauge
  couplings while $\tilde{g}$ represents the `bulk' gauge
  coupling. The left (right) handed light fermions are mostly
  localized at site 0 (2) while their heavy counterparts are mostly at
  site 1. The links connecting sites 0 and 1 and sites 1 and 2 are nonlinear sigma model fields while the one connecting sites 0 and 2 is
  a linear sigma field. Site 2 is dotted to indicate that only the $\tau_3$ component is gauged.}
\label{fig:Triangle}
\end{center}
\end{figure}

The Top Triangle Moose model~\cite{Chivukula:2009ck} is shown in moose
notation in Fig.~\ref{fig:Triangle}.  The circles represent global
SU(2) symmetry groups; the full SU(2) at sites 0 and 1 are gauged with
gauge couplings $g$ and $\tilde g$, respectively, while the $\tau^3$
generator of the global SU(2) at site 2 is gauged with U(1) gauge
coupling $g^{\prime}$.  The lines represent spin-zero link fields
which transform as a fundamental (anti-fundamental) representation of
the group at the tail (head) of the link.  $\Sigma_{01}$ and
$\Sigma_{12}$ are nonlinear sigma model fields, while $\Phi$ (the
top-Higgs doublet) is a linear sigma model field.

The kinetic energy terms of the link fields corresponding to these
charge assignments are:
\begin{equation}
   \mathcal{L}_{gauge}=
   \frac{F^{2}}{4} 
   \textrm{Tr}[(D_{\mu}\Sigma_{01})^{\dagger}D^{\mu}\Sigma_{01}]
   + \frac{F^{2}}{4}
   \textrm{Tr}[(D_{\mu}\Sigma_{12})^{\dagger}D^{\mu}\Sigma_{12}]
   + (D_{\mu}\Phi)^{\dagger}D^{\mu}\Phi,
\label{eqn:Gauge L}
\end{equation}
where the covariant derivatives are:
\begin{eqnarray}
   D_{\mu}\Sigma_{01} &=& \partial_{\mu}\Sigma_{01} + igW_{0\mu} \Sigma_{01}
   - i\tilde{g}\Sigma_{01}W_{1\mu}, \nonumber \\
   D_{\mu}\Sigma_{12} &=& \partial_{\mu}\Sigma_{12} 
   + i\tilde{g}W_{1\mu} \Sigma_{12} 
   - ig^{\prime} \Sigma_{12}\tau^{3}B_{\mu}, \nonumber \\
   D_{\mu}\Phi &=& \partial_{\mu}\Phi + igW_{0\mu} \Phi
   - \frac{ig^{\prime}}{2}B_{\mu}\Phi.
\label{eqn:covariant}
\end{eqnarray}
Here the gauge fields are represented by the matrices $W_{0\mu} =
W_{0\mu}^{a}\tau^{a}$ and $W_{1\mu} = W_{1\mu}^{a}\tau^{a}$, where
$\tau^a = \sigma^a/2$ are the generators of SU(2).  The nonlinear
sigma model fields $\Sigma_{01}$ and $\Sigma_{12}$ are 2$\times$2
special unitary matrix fields.  To mimic the symmetry breaking caused by
underlying technicolor and topcolor dynamics, we assume all link
fields develop vacuum expectation values (vevs):
\begin{equation}
   \langle \Sigma_{01} \rangle = \langle \Sigma_{12} \rangle 
   = \mathbf{1}_{2\times 2}, \qquad \qquad
   \langle \Phi \rangle = 
   \left( \begin{array}{c} f/\sqrt{2} \\ 0 \end{array} \right).
\end{equation}
In order to obtain the correct amplitude for muon decay, we parameterize the
vevs in terms of a new parameter $\omega$,
\begin{equation}
   F=\sqrt{2}\,v\,\cos\,\omega, \qquad \qquad 
   f=v\,\sin\,\omega,
   \label{eq:Ff}
\end{equation}
where $v = 246$~GeV is the weak scale.  As a consequence of the
vacuum expectation values, the gauge symmetry is broken all the way
down to electromagnetism and we are left with massive gauge bosons
(analogous to techni-resonances), top-pions and a top-Higgs. To keep track of
how the degrees of freedom are partitioned after we impose the
symmetry breaking, we expand $\Sigma_{01}$, $\Sigma_{12}$ and $\Phi$
around their vevs. The coset degrees of freedom in the bi-fundamental
link fields $\Sigma_{01}$ and $\Sigma_{12}$ can be described by
nonlinear sigma fields:
\begin{equation}
   \Sigma_{01}=\textrm{exp}\left(\frac{2i\pi_{0}^{a}\tau^{a}}{F}\right),
   \qquad \qquad 
   \Sigma_{12}=\textrm{exp}\left(\frac{2i\pi_{1}^{a}\tau^{a}}{F}\right),
\label{eq:vevs}
\end{equation}
while the degrees of freedom in $\Phi$ fill out a linear
representation,
\begin{equation}
   \Phi= \left( \begin{array}{c}
     (f + H_t + i\pi_{t}^{0})/\sqrt{2} \\
     i\pi^{-}_{t} \end{array} \right).
\label{eqn:phi representation}
\end{equation}

The gauge-kinetic terms in Eq.~\eqref{eqn:Gauge L} yield mass matrices
for the charged and neutral gauge bosons. 
The photon remains massless and is given by the exact expression
\begin{equation}
  A_{\mu} = \frac{e}{g} W_{0 \mu}^3 + \frac{e}{\tilde g} W_{1 \mu}^3
  + \frac{e}{g^{\prime}} B_{\mu},
\end{equation}
where $e$ is the electromagnetic coupling. Normalizing the photon eigenvector, we get the relation between the coupling constants:
\begin{equation}
 \frac{1}{e^2}=\frac{1}{g^2}+\frac{1}{\tilde{g}^2}+\frac{1}{g'^2}.
\end{equation}
This invites us to conveniently parametrize the gauge couplings in terms of $e$ by

\begin{equation}
  g = \frac{e}{\sin\theta \cos\phi} = \frac{g_0}{\cos\phi}, \qquad \qquad
  \tilde g = \frac{e}{\sin\theta \sin\phi} = \frac{g_0}{\sin\phi}, 
  \qquad \qquad
   g^{\prime} = \frac{e}{\cos\theta}.
\label{eqn:gauge couplings}
\end{equation}
We will take $\tilde g \gg g$, which implies that $\sin\phi \equiv x$
is a small parameter. The result of the diagonalization of the gauge boson matrices perturbatively in $x$ is summarized in Appendix A.

Counting the number of degrees of freedom, we see that there are six
scalar degrees of freedom on the technicolor side ($\Sigma_{01},
\Sigma_{12}$) and four on the topcolor side ($\Phi$). Six of these
will be eaten to form the longitudinal components of the $W^{\pm}$,
$Z^0$, $W^{\prime \pm}$, and $Z^{\prime 0}$.  This leaves one isospin
triplet of scalars and the top-Higgs $H_t$ as physical states in the
spectrum.  While the interactions in Eq.~\eqref{eqn:Gauge L} are
sufficient to give mass to the gauge bosons, the top-pions and
top-Higgs remain massless at tree level. Quantum corrections will give
the top-pions a mass, however this loop-level mass is far too small to be consistent with experimental constraints. To generate phenomenologically
acceptable masses for the top-pions and top-Higgs, we add two
additional interactions:
\begin{equation}
   \mathcal L_M =  - \lambda\, {\rm Tr} 
   \left( M^{\dagger}M - \frac{f^2}{2} \right)^2 
   - \kappa f^2\, {\rm Tr} 
   \left| M - \frac{f}{\sqrt 2} \Sigma_{01}\Sigma_{12} \right|^2,
\label{eq:Vhiggs}
\end{equation}
where the first of these interactions arises from topcolor interactions, and the second from ETC-like interactions~\cite{Eichten:1979ah}. Here $\lambda$ and $\kappa$ are two new parameters, $f$ is the same
vacuum expectation value appearing in Eq.~\eqref{eq:Ff}, and $M$ is
the $\Phi$ field expressed as a matrix, schematically given by 
$M = (\tilde \Phi, \Phi)$ with $\tilde \Phi = -i \sigma_2 \Phi^*$:
\begin{equation}
   M = \left( \begin{array}{cc}
     i \pi_t^+ & (f + H_t + i \pi_t^0)/\sqrt{2} \\
     (f + H_t - i \pi_t^0)/\sqrt{2} & i \pi_t^- \end{array} \right),
\end{equation}
where $\pi_t^+ = (\pi_t^-)^*$.  The first term in
Eq.~\eqref{eq:Vhiggs} depends only on the modulus of $M$, and
therefore contributes only to the mass of the top-Higgs. The second
term gives mass to both the top-Higgs and the physical (uneaten)
combination of pion fields, as we will show shortly.  Because these
masses depend on two parameters, $\lambda$ and $\kappa$, we can treat
the mass of the top-Higgs and the common mass of the uneaten top-pions
as two independent parameters. In addition to generating masses, the
potential in Eq.~\eqref{eq:Vhiggs} also induces interactions between
the top-Higgs and top-pions which can be important.

Finally, we note that the mass terms for the light fermions arise from Yukawa couplings of the fermionic fields with the nonlinear sigma fields
\begin{eqnarray}
\mathcal{L} & = & M_{D}\left[\epsilon_{L}\bar{\psi}_{L0}\Sigma_{01}\psi_{R1}+\bar{\psi}_{R1}\psi_{L1}+\bar{\psi}_{L1}\Sigma_{12}\left(\begin{array}{cc}
\epsilon_{uR} & 0\\
0 & \epsilon_{dR}\end{array}\right)\left(\begin{array}{c}
u_{R2}\\
d_{R2}\end{array}\right)\right].
\label{eqn:Light fermion mass}
\end{eqnarray}
We have denoted the Dirac mass that sets the scale of the heavy fermion masses as $M_D$.  Here, $\epsilon_{L}$ is a parameter that describes the degree of delocalization of the left handed fermions and is flavor universal. All the flavor violation for the light fermions is encoded in the last term; the delocalization parameters for the right handed fermions, $\epsilon_{fR}$, can be adjusted to realize the masses and mixings of the up and down type fermions.  The mass of the top quark arises from similar terms with a unique left-handed delocalization parameter $\epsilon_{tL}$ and also from a unique Lagrangian term reflecting the coupling of the top-Higgs to the top quark:
\begin{equation}
\mathcal{L}_{top}=-\lambda_{t}\bar{\psi}_{L0}\,\Phi\, t_{R}+h.c.\label{eq:top quark mass L}
\end{equation}
Details of the fermion masses and mass eigenstates are given in Appendix A.

\section{Physical top-pions and their couplings}
\label{sec:couplings}

The next step towards understanding top-pion phenomenology is to
identify the combination of degrees of freedom which make up the
physical (uneaten) top-pions.  While the top-Higgs $H_t$ remains a
mass eigenstate, the pions $\pi_0^a$, $\pi_1^a$ and $\pi_t^a$ mix.  We
can identify the physical top-pions as the linear combination of
states that cannot be gauged away. We do this by isolating the
Goldstone boson states that participate in interactions of the form
$V_{\mu} \partial^{\mu} \phi$ in the Lagrangian.

We start by expanding the nonlinear sigma fields to first order in
$\pi/F$,
\begin{eqnarray}
   \Sigma_{01} &=& 1 + \frac{2i\pi_{0}^{a}\tau^{a}}{F} 
   + \mathcal{O}\left(\frac{\pi^2}{F^2}\right),
   \label{eqn:sigma01} \\
   \Sigma_{12} &=& 1 + \frac{2i\pi_{1}^{a}\tau^{a}}{F}
   + \mathcal{O}\left(\frac{\pi^2}{F^2}\right).
\label{eqn:sigma12} 
\end{eqnarray} 
Plugging this in Eq.~\eqref{eqn:Gauge
  L}\,\footnote{Here and in the Appendices, the subscripts appearing in
  the fields will refer to the ``site'' numbers and the superscripts
  will be reserved for $SU(2)$ indices.}, we can read off the various interaction terms. The complete details are given in Appendix B. Here, we concentrate on the gauge-Goldstone mixing terms:
\begin{equation}
   \mathcal{L}_{\textrm{mixing}} =
   \frac{g}{2} W_0^{a\mu} \partial_{\mu} \left[F\pi_0^a + f\pi_t^a \right] 
   + \frac{\tilde{g}}{2} W_1^{a\mu} \partial_{\mu} 
   \left[F \pi_1^a - F \pi_0^a \right]
   - \frac{g^{\prime}}{2} B_2^{\mu} \partial_{\mu} 
   \left[F \pi_1^3 + f \pi_t^3 \right].
   \label{eq:gauge-Goldstone}
\end{equation}
Note that the pion combination in the third term can be written as a linear
combination of those appearing in the first two terms:
\begin{equation}
  F \pi_1^3 + f \pi_t^3 = [F \pi_0^3 + f \pi_t^3] + [F \pi_1^3 - F \pi_0^3].
\end{equation}
The two eaten triplets of pions span the linear combinations that
appear in the first two terms of Eq.~\eqref{eq:gauge-Goldstone},
leaving the third linear combination as the remaining physical
top-pions, which we will denote $\Pi_t^a$:
\begin{equation}
   \Pi_t^a = - \sin\omega \left(\frac{\pi_0^a + \pi_1^a}{\sqrt{2}} \right)
   + \cos\omega \, \pi_t^a,
\end{equation}
where we have normalized the state properly using the definitions
of $F$ and $f$ in Eq.~\eqref{eq:Ff}.

The physical top-pions can also be identified by expanding the top-Higgs
potential given in Eq.~\eqref{eq:Vhiggs} and collecting the mass terms.
The masses of $H_t$ and $\Pi_t^a$ are given by,
\begin{equation}
   M^2_H = 2 v^2 (\kappa + 4 \lambda) \sin^2{\omega},
   \qquad \qquad M^2_{\Pi_t} = 2 v^2 \kappa \tan^2{\omega},
\label{eq:pimass}
\end{equation}
while the other two linear combinations of pions are massless, as true Goldstone bosons should be.  Equation~(\ref{eq:Vhiggs}) also
contains trilinear couplings between $H_t$ and two top-pions; we find
\begin{eqnarray}
  \lambda_{H_t \Pi_t^0 \Pi_t^0} = \lambda_{H_t \Pi_t^+ \Pi_t^-}
  &=& 2 v \sin{\omega} \left( \kappa \sin^2{\omega} \tan^2{\omega} 
  + 4 \lambda \cos^2{\omega} \right) \nonumber \\
  &=& \frac{1}{2v \sin{\omega}} \left[(M^2_H - 2 M^2_{\Pi_t})\cos{2\omega} 
    + M^2_H \right].
\end{eqnarray}
These couplings are important for top-Higgs decays when $M_H > 2
M_{\Pi_t}$.  

Having worked out the physical top-pion combination, all that remains
is to express the interactions in the Lagrangian in terms of mass
eigenstates. The top-pion combination is given above, while the gauge
boson and fermion mass eigenstates are given in
Ref.~\cite{Chivukula:2009ck} and are summarized in Appendix A. This conversion is straight-forward,
but tedious, so we just summarize the results for the three-point
couplings in Tables~\ref{tab:pipiV}--\ref{tab:H-couplings}.  We write
the couplings in terms of $x$,  $\sin\theta$, and $g_{0}$, with the latter two defined as in Eq.~\eqref{eqn:gauge couplings}.  The results in this section are given as
an expansion in powers of $x$ and include terms up to order $x^2$.

\begin{table}
\begin{center}
\begin{tabular}{cc}
\hline \hline
Vertex & Strength  \\ \hline
$A_{\mu}\Pi_t^+\Pi_t^-$ & $e\,(p_{\Pi^+}-p_{\Pi^-})_{\mu}$ \\
$Z_{\mu}\Pi_t^+\Pi_t^-$ & $\frac{g_0}{\cos\theta} 
   \left[\left(\frac{1}{2} - \sin^2\theta \right)
   + \frac{x^2}{16} \sec^2\theta
   \left(2 + \cos2\theta \,\sec^2\omega \right) \right]
   (p_{\Pi^-}-p_{\Pi^+})_{\mu}$ \\ 
$Z^{\prime}_{\mu}\Pi_t^+\Pi_t^-$ & $\frac{g_0}{2x} \left[\sin^2\omega
   - \frac{x^2}{16}(7+\cos 2\omega) \sec^2\theta \right] 
  (p_{\Pi^-}-p_{\Pi^+})_{\mu}$ \\ 
$W_{\mu}^{+}\Pi_t^0\Pi_t^-$ & $-\frac{g_{0}}{2} \left[1
   + \frac{x^{2}}{8}(2+\cos 2\omega) \sec^2\omega \right]
   (p_{\Pi^0}-p_{\Pi^-})_{\mu}$ \\ 
$W_{\mu}^{\prime +}\Pi_t^0\Pi_t^-$ & $-\frac{g_0}{2x}
   \left[\sin^2\omega
   - \frac{x^2}{16}\left(7+\cos 2\omega \right) \right]
   (p_{\Pi^0}-p_{\Pi^-})_{\mu}$ \\ 
\hline \hline
\end{tabular}
\caption{\label{tab:pipiV} Couplings of two top-pions to a vector boson.
These have been calculated to $\mathcal{O}(x^{2})$.  Here $p_{\Pi}$ is the
outgoing momentum of particle $\Pi$.}
\end{center}
\end{table}

\begin{table}
\begin{center}
\begin{tabular}{cc}
\hline \hline
Vertex & Strength  \\ \hline
$\Pi_t^-A_{\mu}W_{\nu}^+$ & $0$ \\
$\Pi_t^-A_{\mu}W^{\prime +}_{\nu}$ & $0$ \\
$\Pi_t^-Z_{\mu}W_{\nu}^+$ & $-\frac{i e^2 x^2}{16}
   v\,\sec^3\theta\,\tan\,\omega$ \\ 
$\Pi_t^-Z_{\mu}W_{\nu}^{\prime +}$ & $-\frac{i\,g_0^2}{4x} 
   v\,\sec\theta\,\sin2\omega 
   \left[1 + \frac{x^2}{16}\left(3+5\,\cos2\theta \right)
     \sec^2\theta \right]$ \\ 
$\Pi_t^-Z^{\prime}_{\mu}W^+_{\nu}$ & $\frac{i\,g_0^2}{4x} v \sin2\omega
   \left[1
   + \frac{x^2}{16}\left(5+3\,\cos2\theta \right)\sec^2\theta \right]$ \\
$\Pi_t^-Z^{\prime}_{\mu}W^{\prime +}_{\nu}$ & $-\frac{i\,g_0^2}{8} v \,
   \tan^2\theta \left[\sin 2\omega
   + \frac{x^2}{32}\left(10-2\,\cos 2\theta
   + 3\,\cos [2(\theta-\omega)] + 2\,\cos 2\omega
   + 3\,\cos [2(\theta+\omega)] \right) \sec^2 \theta\,
   \tan\omega \right]$ \\ 
$\Pi_t^0W^+_{\mu}W^{\prime -}_{\nu}$ & $\frac{ig_0^2}{4x} v \,
   \sin 2\omega \left(1 + \frac{x^2}{2}\right)$ \\ 
\hline \hline
\end{tabular}
\caption{\label{tab:piVV} Couplings of a top-pion to a pair of gauge
  bosons.  The corresponding Feynman rule is obtained by inserting a
  $g_{\mu\nu}$ in each coupling.  These have been calculated to
  $\mathcal{O}(x^{2})$.}
\end{center}
\end{table}

\begin{table}
\begin{center}
\begin{tabular}{cc}
\hline \hline
Vertex & Strength \\
\hline
$H_tW^+_{\mu}W^-_{\nu}$ & $\frac{g_{0}^{2}}{2} v \sin\omega 
   \left(1+\frac{3\,x^2}{4} \right)$ \\ 
$H_tW^{\prime +}_{\mu}W^-_{\nu}$ & $-\frac{g_{0}^{2} x}{4} v \sin\omega$ \\ 
$H_tW^{\prime +}_{\mu}W^{\prime -}_{\nu}$ & $\frac{g_{0}^{2} x^{2}}{8} v
   \sin\omega$ \\ 
$H_tZ_{\mu}Z_{\nu}$ & $\frac{g_0^2}{4 \cos^2\theta} v
   \sin\omega \left[1 + \frac{x^2}{4}\left(1 + 2\cos 2\theta \right)
   \sec^2 \theta \right]$ \\ 
$H_tZ^{\prime}_{\mu}Z_{\nu}$ & $-\frac{g_{0}^{2} x}{4 \cos^2\theta} v 
   \sec\theta \cos 2\theta \sin\omega$ \\ 
$H_tZ^{\prime}_{\mu}Z^{\prime}_{\nu}$ & $\frac{g_{0}^{2} x^2}{16 \cos^2\theta} 
   v \sec^2\theta \cos^2 2\theta \sin\omega$ \\ 
$H_t \Pi_t^- W^+_{\mu}$ & $\frac{g_0}{2} \cos\omega 
   \left(1+\frac{3 x^2}{8} \right)(p_H - p_{\Pi^-})_{\mu}$ \\ 
$H_t \Pi_t^- W^{\prime +}_{\mu}$ & $-\frac{g_{0} x}{4} \cos\omega
   (p_H - p_{\Pi^-})_{\mu}$ \\ 
$H_t \Pi_t^0 Z_{\mu}$ & $-\frac{g_{0}}{2 \cos\theta} \cos\omega 
   \left[1 + \frac{x^2}{8} \left(1 + 2\cos 2\theta \right) \sec^2\theta 
   \right](p_H - p_{\Pi^0})_{\mu}$ \\ 
$H_t \Pi_t^0 Z^{\prime}_{\mu}$ & $\frac{g_{0} x}{4 \cos\theta} \cos\omega 
   \sec\theta \cos 2\theta (p_H - p_{\Pi^0})_{\mu}$ \\ 
\hline \hline
\end{tabular}
\end{center}
\caption{\label{tab:H-couplings} Three-point couplings of the
  top-Higgs, again calculated to $\mathcal{O}(x^2)$.  The Feynman rules
  for the couplings involving two gauge bosons are obtained by
  multiplying the coupling strength given here by $i g_{\mu\nu}$.}
\end{table}

Notice in particular that the couplings of the heavy gauge bosons
$Z^{\prime}$ and $W^{\prime \pm}$ to two top-pions are proportional to
the large gauge coupling $\tilde g = g_0/x$ associated with site 1.  The leading term in
these couplings is in fact $\tilde g \sin^2\omega/2$, with the two
$\sin\omega$ factors reflecting the overlap of the $\Pi_t^a$
wavefunction with the combination of nonlinear sigma fields $(\pi_0^a
+ \pi_1^a)/\sqrt{2}$. 
The couplings of the top-Higgs and top pions to third generation
fermions (and their heavy partners) can be likewise computed, by
plugging in the mass eigenstates into the top quark mass term, Eq.~\eqref{eq:top quark mass L}, with $\Phi$ is given by Eq.~\eqref{eqn:phi representation}). The results are shown in Table IV, written in terms of the parameter $a= v\,\textrm{sin}\,\omega/ \sqrt{2}M_{D}$.  

We have also worked out the four-point interactions. While these are less important phenomenologically, we list the mass-basis couplings in  Appendix C for completeness.

\begin{table}
\begin{center}
\renewcommand{\arraystretch}{2.2}
  \begin{tabular}{| c || c | }
    \hline
     Vertex & Strength  \\ \hline
     $H_t\,t_{L}\bar{t}_R+h.c$ & $\frac{\lambda_t}{\sqrt{2}}\left[-1+\frac{(1+a^2)(x^2+2 \epsilon_{tR}^2)+4\sqrt{2}ax\, \epsilon_{tR}}{4(a^2-1)^2} \right]$ \\ \hline
     $H_t\,T_{L}\bar{T}_R+h.c$ & $-\frac{\lambda_t}{\sqrt{2}}\left[\frac{a (x^2+2 \epsilon_{tR}^2)+\sqrt{2}(1+a^2)x\,\epsilon_{tR}}{2(a^2-1)^2} \right]$ \\ \hline
     $H_t\,t_{L}\bar{T}_R+h.c$ & $\lambda_t \left[\frac{a\,x+\sqrt{2}\epsilon_{tR}}{2(a^2-1)} \right]$ \\ \hline
     $H_t\,t_{R}\bar{T}_L+h.c$ & $\lambda_t \left[\frac{x+\sqrt{2}a\,\epsilon_{tR}}{2(a^2-1)} \right]$ \\ \hline
     $\Pi_{t}^0\bar{t}_Lt_R-h.c.$ & $\frac{i\lambda_t}{\sqrt{2}}\left[-\textrm{cos}\,\omega + \frac{(a^2+\textrm{cos}2\omega)\textrm{sec}\,\omega(x^2+2\,\epsilon_{tR}^2)+\frac{2\sqrt{2}}{a}x\,\epsilon_{tR}\left[2a^2\textrm{cos}\,\omega+(a^2-1)\textrm{sin}\,\omega\,\textrm{tan}\,\omega \right]}{4\sqrt{2}(a^2-1)^2} \right]$ \\ \hline
     $\Pi_{t}^0\bar{T}_LT_R-h.c.$ & $\frac{i\lambda_t}{4\sqrt{2}}\left[\frac{ 2\sqrt{2}\,x\,\epsilon_{tR}(a^2+\textrm{cos}2\omega)+(2a^2\textrm{cos}\,\omega+(a^2-1)\textrm{sin}\,\omega\,\textrm{tan}\,\omega)\,(x^2+2\,\epsilon_{tR}^2)}{(a^2-1)^2} \right]$ \\ \hline
     $\Pi_{t}^0\bar{t}_LT_R-h.c.$ & $\frac{i\lambda_t}{(a^2-1)}\left[\frac{x\textrm{sec}\,\omega(-1+3a^2+(1+a^2)\,\textrm{cos}2\omega)}{8a}+\frac{\epsilon_{tR}\textrm{cos}\,\omega}{\sqrt{2}} \right]$ \\ \hline
     $\Pi_{t}^0\bar{T}_Lt_R-h.c.$ & $\frac{i\lambda_t}{(a^2-1)}\left[\frac{x\,\textrm{cos}\,\omega}{2}+\frac{\epsilon_{tR}\,\textrm{sec}\,\omega(-1+3a^2+(1+a^2)\,\textrm{cos}2\omega)}{4\sqrt{2}a} \right]$ \\ \hline
     $\Pi_{t}^{-}t_R\bar{b}_L-h.c.$ & $i\lambda_t\left[\textrm{cos}\,\omega-\frac{x^2\left(a^4+(a^4-2a^2+2)\,\textrm{cos}2\omega \right)\textrm{cos}\,\omega}{8(a^2-1)^2}-\frac{x\,\epsilon_{tR}\,\textrm{sec}\,\omega \left(-2+5a^2-a^4+(a^4-a^2+2)\,\textrm{cos}2\omega \right)\textrm{cos}\,\omega}{4\sqrt{2}a(a^2-1)^2}-\frac{\epsilon_{tR}^2\,\textrm{cos}\,\omega}{2(a^2-1)^2} \right]$ \\ \hline
     $\Pi_{t}^{-}T_R\bar{B}_L-h.c.$ & $\frac{i\lambda_t\,\textrm{sec}\,\omega}{(a^2-1)}\left[-\frac{x^2\,(-1+3a^2+(1+a^2)\,\textrm{cos}2\omega)}{8a}-\frac{x\,\epsilon_{tR}(1+3\textrm{cos}2\omega)}{4\sqrt{2}a}+\frac{\textrm{sin}^2\omega\,\epsilon_{tR}^2}{2a} \right]$ \\ \hline
     $\Pi_{t}^{-}T_R\bar{b}_L-h.c.$ & $-\frac{i\lambda_t}{(a^2-1)}\left[\frac{x\left(-1+3a^2+(1+a^2)\,\textrm{cos}2\omega \right)\,\textrm{sec}\,\omega}{4\sqrt{2}a}+\epsilon_{tR}\,\textrm{cos}\,\omega \right]$ \\ \hline
     $\Pi_{t}^{-}t_R\bar{B}_L-h.c.$ & $i\lambda_t\left[\frac{x\,\textrm{cos}\,\omega}{\sqrt{2}}-\frac{\epsilon_{tR}\,\textrm{sin}\,\omega\,\textrm{tan}\,\omega}{2a} \right]$ \\ \hline
    
    \hline
  \end{tabular}

  \caption{The couplings of the top-Higgs and top pions to third generation fermions and their heavy partners, calculated to $\mathcal{O}(x^{2},\epsilon_{tR}^2)$.}
\end{center}
\label{tab:H-fermion}
\end{table}

\section{Top-Higgs and top-pion phenomenology}
\label{sec:pheno}

 We are now prepared to investigate the phenomenology of the new states related to the top quark: the top-Higgs $(H_t)$, the top-pions $(\Pi_t)$, and the heavy vector fermion partner of the top quark $(T)$.  First, we will show how existing Tevatron data can be applied to place limits on the top triangle moose model.  Essentially, rescaling to take altered coupling values into account allows  limits derived for other models to be transformed into limits on our model's top-pions and top-Higgs. Next, we study top-Higgs and top-pion production at LHC.  As indicated in Figure~(\ref{fig:higgs_prod}) below, the new scalars can be produced either directly, through gluon fusion via a top loop, or indirectly, via decays of the heavy $T$ quarks. 
\begin{figure}[!ht]
\includegraphics[scale=1.0]{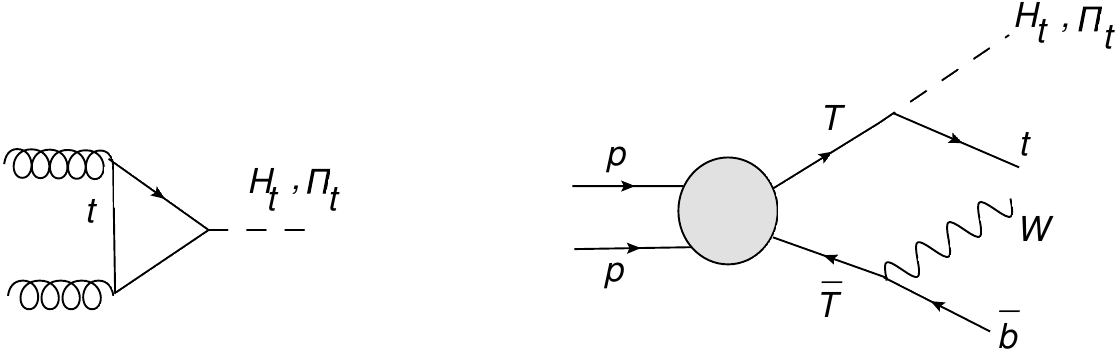}
\caption{\label{fig:feynman1} Feynman diagrams for the single production of the top-Higgs or top pion. Direct production proceeds  via a top loop (left).  Indirect production occurs as a decay product of heavy T quarks (right), where the pair production of the T quarks proceeds via gluon fusion and quark annihilation.}
\label{fig:higgs_prod}
\end{figure}
 
 \noindent The multiple production modes will make it possible to confirm that the scalar one has discovered is, in fact, the $H_t$ of this model, rather than some other scalar state.  We examine the branching ratios of the $H_t, \Pi_t$, and $T$, in order to identify production channels at LHC that are likely to lead to discovery of these new particles.  Then, we discuss how already-planned searches at LHC can be repurposed or extended to yield information about the individual states and the relationships between them.

\subsection{Current constraints on parameter values}
Before starting our phenomenological analysis, let us briefly recall some of the limits on the different parameters of our model.  First, within the gauge sector there is the mass of the $W'$ boson, $M_{W'}$, and the ratio of gauge couplings $x$.  Second, there are parameters related to fermion masses: the quantities $\epsilon_L$, $\epsilon_{fR}$, $\epsilon_{tL}$, $\epsilon_{tR}$, and $M_D$.  In addition, from the top-Higgs sector, we have $\lambda_t$, $\textrm{sin}\,\omega$ and the masses of the top-pion and the top-Higgs, $M_{\Pi_t}$ and $M_{H_t}$.

As discussed in \cite{Chivukula:2006cg}, the value of the $W'$ mass is constrained to lie above 380 GeV by the LEP II measurement of the triple gauge boson vertex  and to lie below 1.2 TeV by the need to maintain perturbative unitarity in $W_L W_L$ scattering\protect{\footnote{Strictly speaking, the upper bound on $M_{W'}$ in this model should be slightly different than the one in the three site model, because the formula for $M_{W'}$ is slightly different. However, to the extent that $f \ll F$ ($\sin\omega \ll 1$), this may be neglected.}}.  We will use the illustrative $M_{W'} = 500$ GeV in our calculations (except where noted otherwise) both for definiteness and because the value of the parameter $\epsilon_L$ is then derivable from $M_{W'}$  (and $x$) via ideal delocalization, as shown below.

In principle, the values of the various $\epsilon_{fR}$ are proportional to the masses of the light fermions; since we will be working in the limit $m_f \rightarrow 0$ for fermions other than the top quark, we will set $\epsilon_{fR}=0$.  Similarly, since the top quark mass depends very little on $\epsilon_{tR}$ we will set $\epsilon_{tR}=0$ as well for simplicity.  In this limit $M_D$ corresponds to the (degenerate) masses of the heavy fermionic partners of the light ordinary fermions, and is closely related (as shown in Appendix A) to the mass of the heavy partner of the top quark.  We will set $M_D$ to the illustrative value $M_D = 400$ GeV in calculations not depending strongly on the precise value, and will otherwise show how results vary with $M_D$.

Within the top-pion sector we will set $M_{H_t}$ and $M_{\Pi_t}$ to the illustrative values $M_{H_t} = 250$ GeV and $M_{\Pi_t} = 200$ GeV when the precise value is not critical and will otherwise show how results vary with the values of these masses.  Likewise, we will allow $\sin\omega$ to vary to show how various physical quantities depend on it; when the dependence is not critical, we will tend to use the illustrative value $\sin\omega = 0.5$.

The remaining parameters, $x$ and the top Yukawa coupling $\lambda_t$, are now derived from the quantities above via:
\begin{eqnarray}
x &=& \sqrt{2} \, \epsilon_L = \frac{2 \cos\omega M_W}{M_{W^{\prime}}} \nonumber \\
\lambda_t &=& \frac{\sqrt{2} \, m_t}{v \sin\omega}
\left[ \frac{M_D^2 (\epsilon_L^2 + 1) - m_t^2}{M_D^2 - m_t^2} \right],
\label{eq:lambdat}
\end{eqnarray}
where $m_t$ is the physical top quark mass and in the last expression we have set $\epsilon_{tR} = 0$.  The relationship between $\epsilon_L$ and $x$ is imposed by ideal delocalization; similarly, as discussed below in subsection IV.A.3, flavor constraints tend to force $\epsilon_{tL} \simeq \epsilon_{L}$, so the value of this last parameter is set as well.

\subsubsection{Tevatron limits from Higgs searches}
\label{sec:Higgs Limits}

The Tevatron experiments analyzed the channel $gg \to H \to WW$ and set upper bounds on the cross-section as a function of $M_H$ in Ref.~\cite{Aaltonen:2010sv}. We can adapt this data to our model by appropriately rescaling the couplings involved in the following way: Because of the $\sin\omega$ factor in the denominator of Eq.~\eqref{eq:lambdat} for $\lambda_t$ above, couplings of $H_t$ to top quarks are enhanced compared to those of the SM Higgs, particularly for small $\sin\omega$.  This leads to an enhanced cross section for $H_t$ production in gluon fusion, scaling proportional to $(\lambda_t/\lambda_t^{\rm SM})^2$.  Simultaneously, due to the absence of the decay mode $H_t \to b \bar b$ at low $H_t$ masses, the branching ratio for $H_t \to WW$ is larger than for the SM Higgs for masses below about 160~GeV.  These two features lead to an enhancement of the predicted rate for $gg \to H_t \to WW$ compared to the corresponding SM process, which is already constrained by Tevatron data.

\begin{figure}[h!]
\includegraphics[scale=0.5]{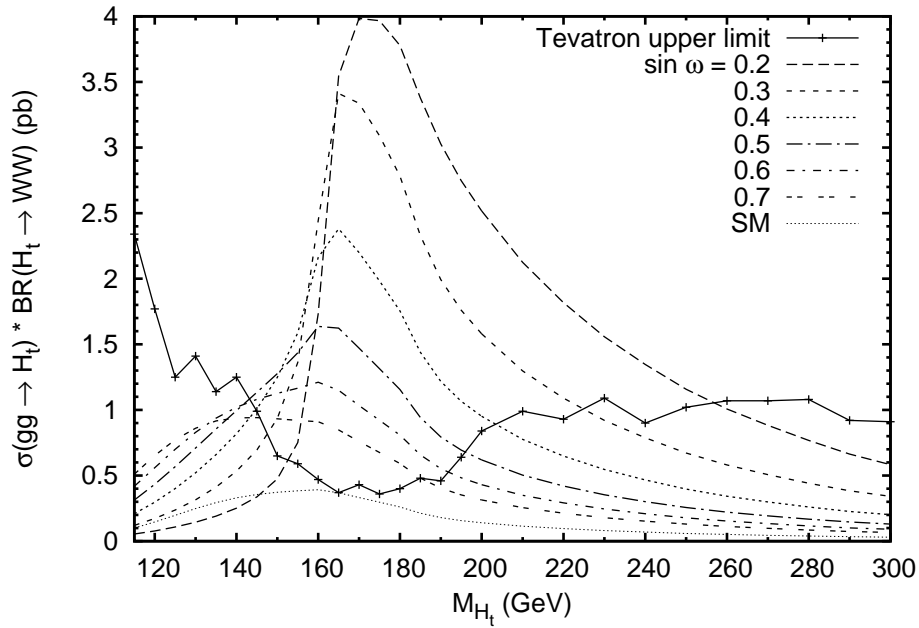}
\caption{Constraints on $\sigma\cdot BR$ for top-Higgs production at the Tevatron as a function of $M_{H_t}$ and $\sin\omega$.  One can clearly see how $\sin\omega < 1$  enhances the production cross-section.  For a given value of $\sin\omega$, the range of masses for which the corresponding curve lies above the ``Tevatron upper limit" curve is excluded.  Each curve for fixed $\sin\omega$ falls off at small $M_{H_t}$ due to a decline in $BR(H_t \to WW)$ and drops off at high $M_{H_t}$ because the falling gluon pdf reduces the production cross-section $\sigma(gg \to H_t)$.}
\label{fig:limit-sin.pdf}
\end{figure}

We can now translate the Tevatron bounds on the cross-section \cite{Aaltonen:2010sv} into constraints on the $\sin\omega$--$M_{H_t}$ parameter space as follows.  We compute the cross section for $gg \to H_t \to WW$ according to the approximation
\begin{align}
	\sigma(gg \to H_t \to WW) &= \sigma^{\rm SM}(gg \to H) 
		\frac{\Gamma(H_t \to gg)}{\Gamma^{\rm SM}(H \to gg)}
		{\rm BR}(H_t \to WW) \\ \nonumber
                &=\sigma^{\rm SM}(gg \to H) \frac{BR( H_t \to gg)\Gamma(H_t)}{BR( H_{\textrm{SM}} \to gg)\Gamma(H_{\textrm{SM}})}{\rm BR}(H_t \to WW),
\end{align}
where $\Gamma^{\rm SM}(H \to gg)$ is the SM partial width of $H$ to gluons computed using HDECAY~\cite{HDECAY}, $\Gamma(H_t \to gg)$ and BR($H_t \to WW$) are the partial width of $H_t$ to gluons and the branching ratio of $H_t$ to $WW$, respectively, computed using our modified version of HDECAY, and $\sigma^{\rm SM}(gg \to H)$ is the SM Higgs gluon fusion cross section, which we take from Table~2 of Ref.~\cite{Anastasiou:2008tj} for $M_H\leq 200$ GeV and compute using the public code RGHIGGS ~\cite{Ahrens:2008qu,Ahrens:2008nc,Ahrens:2010rs} for $M_H> 200$ GeV. For each value of $\sin\omega$, a specific range of masses for the top-Higgs is excluded by the Tevatron data.  For example, for the illustrative value $\sin\omega = 0.5$, the data implies that the mass range $140\ {\rm GeV} < M_{H_t} < 195\ {\rm GeV}$ is excluded.  We present this in Fig.~\ref{fig:limit-sin.pdf} - as can clearly be seen, the scaling of the Yukawa enhances the production cross-section in our model. As we move to larger $M_{H_t}$ values, $\sigma\, \cdot\, \text{BR}$ declines toward zero as the parton distribution function of the gluon falls rapidly.

Turning to the top-pion, we find that there are more important constraints on the $\Pi_t$ masses than those derived from the Tevatron Higgs search limit. These constraints come from limits on rare top decays, from $Zb\bar{b}$ couplings, and from B-physics. We discuss these in turn below.
\subsubsection{Lower bound on the top-pion mass}

If the charged top-pion $\Pi_t^+$ is lighter than the top quark, it can appear in top decays, $t \to \Pi_t^+ b$.  The Tevatron experiments have searched for this process in the context of two-Higgs-doublet models and set upper bounds of about 10--20\% on the branching fraction of $t \to H^+ b$, with $H^+$ decaying to $\tau \nu$ or $c \bar s$~\cite{Aaltonen:2009ke,Abazov:2009zh} - we can use this to set a lower bound on the top-pion mass. In our model, below the $t \bar b$ threshold $\Pi_t^+$ decays via its mixing with $\pi_0^+$ to lighter SM fermions, with couplings controlled by the fermions' SM Yukawa couplings.  The branching fraction of $\Pi_t^+$ to $\tau\nu$ is therefore about 70\%, with the remainder of decays to $c \bar s$.  The
Tevatron studies can then be applied directly to the top-pion.  The relevant limit is BR($t \to \Pi_t^+ b) \lesssim 0.2$ based on D0 data~\cite{Abazov:2009zh}.

In our model, the branching fraction of $t \to \Pi_t^+ b$ is controlled by the top-pion mass, the pion mixing angle $\omega$, and the coupling $\lambda_t$:
\begin{equation}
  \Gamma(t \to \Pi_t^+b) = \frac{G_F m_t}{8 \sqrt{2} \pi}
    \left[m_t^2 R^2 + \mathcal{O}(m_b^2)\right]
    \left[1 - \frac{M_{\Pi_t^+}^2}{m_t^2} \right]^2,
\label{eqn:gamma_tb}
\end{equation}
with
\begin{equation}
  R \equiv \cos\omega \frac{\lambda_t}{\lambda_t^{\rm SM}}
  = \cot\omega \left[ \frac{M_D^2 (\epsilon_L^2 + 1) - m_t^2}{M_D^2 - m_t^2}
    \right].
\end{equation}
To evaluate this expression numerically, we choose the illustrative values: $\sin\omega = 0.5$, $M_D=400$ GeV. Plugging in these values in Eq.~\eqref{eqn:gamma_tb} leads to the constraint $M_{\Pi_t^+} \gtrsim 150$~GeV from top quark decay\footnote{This bound gets stronger as $\sin\omega$ becomes smaller.}.  Because $\Pi_t^0$ and $\Pi_t^+$ are degenerate, this sets a lower bound on both particles' masses. Having established that $M_{\Pi_t}$ cannot be much lighter than $m_t$, we will assume in the rest of the paper that $M_{\Pi_t} > m_t$, so that decays of $\Pi_t^+ \to t \bar b$ dominate~\footnote{LHC experiments should be able to reduce the upper limit on BR($t \to \Pi_t^+ b$) to about $10^{-2}$~\cite{Aad:2009wy,Baarmand:2006dm}, which would push the lower bound on the top-pion mass above 170~GeV for the parameter point considered here.  However, the studies of the LHC reach have been done only for charged Higgs masses below 150~GeV; for higher masses, off-shell decays to $t^*b$ should also be considered.}.

\subsubsection{$\textrm{R}_{\textrm{b}}$ and non-ideal delocalization of the left-handed top quark}

Next, we consider how data on the $Zb\bar{b}$ coupling constrains the allowed values of $\sin\omega$ and $M_{H_t}$.  
At tree level, the $Z$ coupling to left-handed bottom quarks is generically modified due to the profile of the $b_L$ wavefunction at the three sites, yielding~\cite{Chivukula:2009ck},
\begin{equation}
	g_L^{Zbb} = - \frac{e}{s_Wc_W} 
	\left[ \left(1 + \frac{x^2}{4} - \frac{\epsilon_{tL}^2}{2} \right) T_3
	- Q s_W^2 \right].
\end{equation}
The tree-level shift in $R_b$ can be eliminated by imposing \emph{ideal delocalization} on the left-handed fermions~\cite{Chivukula:2009ck}, which means setting:
\begin{equation}
	\epsilon^2_{tL} = (\epsilon^{ideal}_{tL})^2 \equiv \frac{x^2}{2}~.
\end{equation}
This has the additional benefit of decoupling the SM fermions from the $W^{\prime}$, $Z^{\prime}$ gauge bosons, eliminating a potentially dangerous source of 4-fermion operators.

At one loop, $R_b$ receives additional contributions which we parameterize as $\delta g_L$ according to
\begin{equation}
	g_L^{Zbb} = - \frac{e}{s_Wc_W} 
	\left[ \left( 1 + \frac{(\epsilon^{ideal}_{tL})^2}{2} - \frac{\epsilon_{tL}^2}{2} \right) T_3 
	+ \delta g_L - Q s_W^2 \right].
	\label{eq:glzbb}
\end{equation}
The one-loop corrections come from:
\begin{itemize}
 \item Loops involving $W$, $W^{\prime}$, SM fermions, and/or heavy vector-like fermions - these were computed for the three-site model~\cite{Chivukula:2006cg} in Ref.~\cite{Abe:2009ni}.
\item Loops involving the charged top-pion and at least one vector-like heavy fermion. Note that the couplings of $\Pi_t^-$ to one (two) vector-like heavy fermions are suppressed by one (two) powers of $x$ or $\epsilon_{tR}$.
\item Loops involving the charged top-pion and SM fermions, including contributions from the Goldstone boson eaten by the $W$ in the Top Triangle Moose model, which contains an admixture of the original top-Higgs doublet. These contributions were studied in Ref.~\cite{Burdman:1997pf} for a generic topcolor model based on the calculation of the contribution to $R_b$ in the two-Higgs-doublet model done in Ref.~\cite{Denner:1991ie}.
\end{itemize}
In the Top Triangle Moose model that we consider in this paper, most of the top quark mass comes from the topcolor mechanism and the contribution from $\epsilon_{tR}$ is small, no more than a few GeV.  Therefore the contributions to $\delta g_L$ given by the first two sources will be negligible, and the dominant correction comes from the charged top-pion loops. These give a $\lambda_t^2$--enhanced correction to $R_b$ given by

\begin{equation}
	\delta g_L^{\Pi^-_t} = \frac{m_t^2}{16 \pi^2 v^2} \cot^2 \omega 
	\left[ \frac{R}{R-1} - \frac{R \ln R}{(R-1)^2} \right],
	\label{eq:dglpi}
\end{equation}
where $R \equiv m_t^2/M_{\Pi^-_t}^2$; note that $M_Z$ and $m_b$ have been neglected in the loop calculation relative to $m_t$ and $M_{\Pi_t}$.

We now consider the size of the dominant new-physics correction $\delta g_L^{\Pi_t^-}$ and compare it to the experimental constraints on $R_b$.  We can express the new physics contribution to $R_b$ in terms of $\delta g_L^{\rm new}$ according to \cite{Oliver:2002up},
\begin{equation}
	\delta R_b = 2 R_b (1 - R_b) \frac{g_L}{g_L^2 + g_R^2} \delta g_L^{\rm new},
\label{eq:rbdefnn}	
\end{equation}
where $g_L$ and $g_R$ are the SM $Zbb$ couplings at leading order,
\begin{equation}
	g_L = -\frac{1}{2} + \frac{s_W^2}{3}, \qquad \qquad 
	g_R = \frac{s_W^2}{3}, \qquad \qquad {\rm with} \ \ s_W^2 \simeq 0.23.
\end{equation}
To leading order we can insert the SM prediction for $R_b$~\cite{Amsler:2008zzb},
\begin{equation}
	R_b^{\rm SM} = 0.215 \, 84 \pm 0.000 \, 06\,,
\label{eq:rbsmm}	
\end{equation}
in the right-hand side of Eq.~(\ref{eq:rbdefnn}). This yields the convenient numerical expression,
\begin{equation}
	\delta R_b \simeq -0.774 \, \delta g_L^{\rm new}.
\label{eq:numexprrb}	
\end{equation}

The current experimental value of $R_b$ is~\cite{Amsler:2008zzb},
\begin{equation}
	R_b^{\rm expt} = 0.216 \, 29 \pm 0.000 \, 66.
\end{equation}
Subtracting this from the SM prediction Eq.~(\ref{eq:rbsmm}) gives us a value for the left-hand side of Eq.~(\ref{eq:numexprrb}), yielding a constraint on the new physics contribution,
\begin{equation}
	\delta g_L^{\rm new} = (-5.8 \pm 8.6) \times 10^{-4}.
	\label{eq:deltagLlimits}
\end{equation}
This, in turn, implies a 2$\sigma$ (3$\sigma$) upper bound on $\delta g_L^{\rm new}$ of 
$11.4 \times 10^{-4}$ ($20.0 \times 10^{-4}$).

Now let us see what we can deduce about constraints on the parameter space of our model.  Let us first consider ideal delocalization (i.e., no tree-level contribution to $R_b$ from the distribution of the light fermion wavefunction among the sites).  The coefficient in Eq.~\eqref{eq:dglpi} is numerically,
\begin{equation}
	\frac{m_t^2}{16 \pi^2 v^2} = 32 \times 10^{-4}.
\end{equation}
We take $\sin\omega = 0.5$, which yields $\cot^2 \omega = 3$.  When $M_{\Pi_t} = m_t$ (i.e., $R=1$), the function of $R$ in square brackets in Eq.~\eqref{eq:dglpi} is equal to $1/2$.  At this parameter point we thus have,
\begin{equation}
	\delta g_L^{\Pi_t^-} = 48 \times 10^{-4} \qquad \qquad (\cot^2\omega = 3, \ \ M_{\Pi_t} = m_t),
\end{equation}
which is forbidden.  The function of $R$ in square brackets in Eq.~\eqref{eq:dglpi} falls with increasing $M_{\Pi_t}$.  This allows us to put a lower bound on $M_{\Pi_t}$ assuming ideal fermion delocalization and taking $\sin\omega = 0.5$:
\begin{equation}
	M_{\Pi_t} \gtrsim 760 \ (480) \ {\rm GeV} \qquad {\rm at} \ 2\sigma \ (3\sigma).
\end{equation}

A lighter top-pion can be allowed if we shift the left-handed third generation quarks away from ideal delocalization.  A positive one-loop $\delta g_L^{\rm new}$ can be compensated by choosing a smaller $\epsilon_{tL}$.  At our parameter point we have chosen $M_{W^{\prime}} = 500$~GeV, which (with $\sin\omega = 0.5$) yields,  
\begin{equation}
	x = \frac{2 \cos\omega M_W}{M_{W^{\prime}}} \simeq 0.28, \qquad \qquad {\rm or} \qquad\, (\epsilon_{tL}^{ideal})^2\,\simeq 0.039.
\end{equation}
Returning to Eq.~\eqref{eq:glzbb}, the combined tree-level and one-loop new physics contribution can be eliminated by choosing $\epsilon_{tL}$ to satisfy
\begin{equation}
\frac14 \left(\epsilon^2_{tL} - (\epsilon_{tL}^{ideal})^2 \right) + \delta g_L^{\Pi_t^-} = 0 \,.
\end{equation}
More generally, if we define
\begin{equation}
\Delta \epsilon^2_{tL} = \epsilon^2_{tL} - (\epsilon_{tL}^{ideal})^2
\end{equation}
then we can deduce from Eq.~\eqref{eq:deltagLlimits} that the value of $\epsilon_{tL}$ must satisfy
\begin{equation}
\frac14 \Delta \epsilon^2_{tL} + \delta g_L^{\Pi_t^-} < 11.4 \times 10^{-4}\ \   (20.0 \times 10^{-4})
\end{equation}
in order for the predicted value of $R_b$ to agree with experiment at the $2\sigma$ ($3\sigma$) level.

\begin{figure}[th]
\centering
\includegraphics[width=0.60\textwidth]{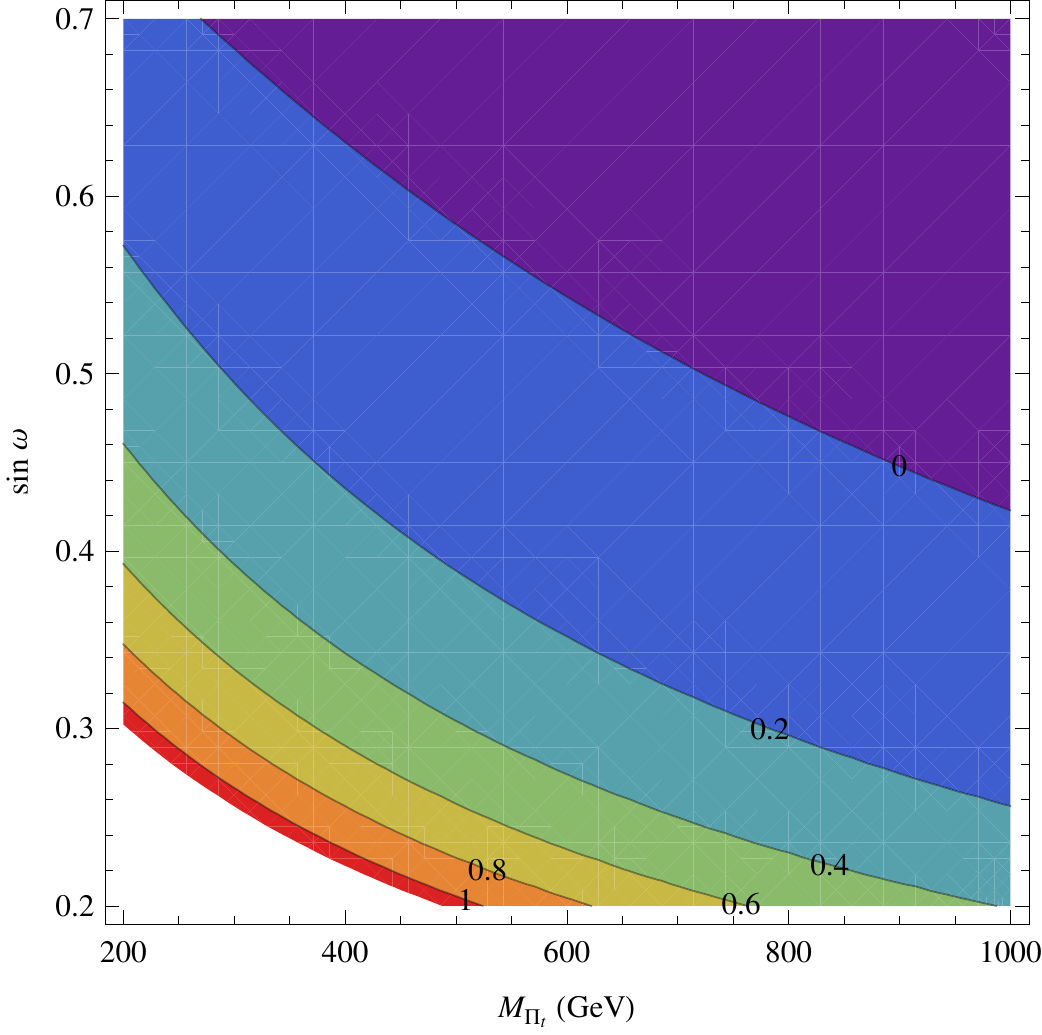}
\caption{A contour plot of the deviation $|\Delta \epsilon^2_{tL}/(\epsilon^{ideal}_{tL})^2|$ required to compensate for the top-pion
contribution to $R_b$, for $M_{W'}=500\, \textrm{GeV}$. The contour boundaries correspond to $|\Delta \epsilon^2_{tL}/(\epsilon^{ideal}_{tL})^2|$
equal to 0 to 1.0 in increments of 0.2. As discussed in the text, for larger values of $M_{W'}$, each contour would correspond to a larger value of $|\Delta \epsilon^2_{tL}/(\epsilon^{ideal}_{tL})^2|$. Note that, for a deviation $|\Delta \epsilon^2_{tL}/(\epsilon^{ideal}_{tL})^2|$ of
order 1, essentially the entire plane of values is allowed.
\label{fig:one}
}
\end{figure}

Figure~\ref{fig:one} shows a contour plot of the fractional deviation  $|\Delta \epsilon^2_{tL}/(\epsilon^{ideal}_{tL})^2|$ from ideal fermion delocalization required in order for top quark delocalization to compensate for top-pion corrections to $R_b$ (meaning agreement at the 90\% CL level).  Note that for a fractional deviation of order 1, essentially the entire $\sin\omega$ vs $M_{\Pi_t}$ plane is allowed. The illustrative value $M_{W'} = 500$ GeV was used in making this plot; since $\epsilon_{tL}^{ideal} \propto M_{W'}^{-1}$, for heavier $W'$ bosons the contours would retain their shape and label but correspond to a  larger value of $\epsilon_{tL}$.

%

Finally, one may worry that changing the value of $\epsilon_{tL}$ from its ideal value might cause problems with flavor changing neutral current constraints. We demonstrate in Appendix D that, in the case of ``next-to-minimal" flavor violation~\cite{Agashe:2005hk}, these do not rule out compensating for the deviation in $R_b$ resulting from top-pion exchange by modifying the delocalization of the third-generation quarks.

\subsection{Top Higgs production and decay}

Having derived the relevant interactions between matter and the top-Higgs/top-pions and understood current constraints on the Top-Triangle moose parameter space, we are ready to move on to phenomenology. In this section and the following, we present the dominant production and decay rates for the top-Higgs and top-pions respectively in a viable region of parameter space. For both the $H_t$ and $\Pi_t$ we consider both direct production $pp \rightarrow H_t,\Pi_t$ and indirect production -- top-Higgses/top-pions which arise from the decays of $T$ quarks.

\subsubsection{Decay Branching Ratios}

\begin{figure}[!ht]
\centering
\includegraphics[width=3.39in]{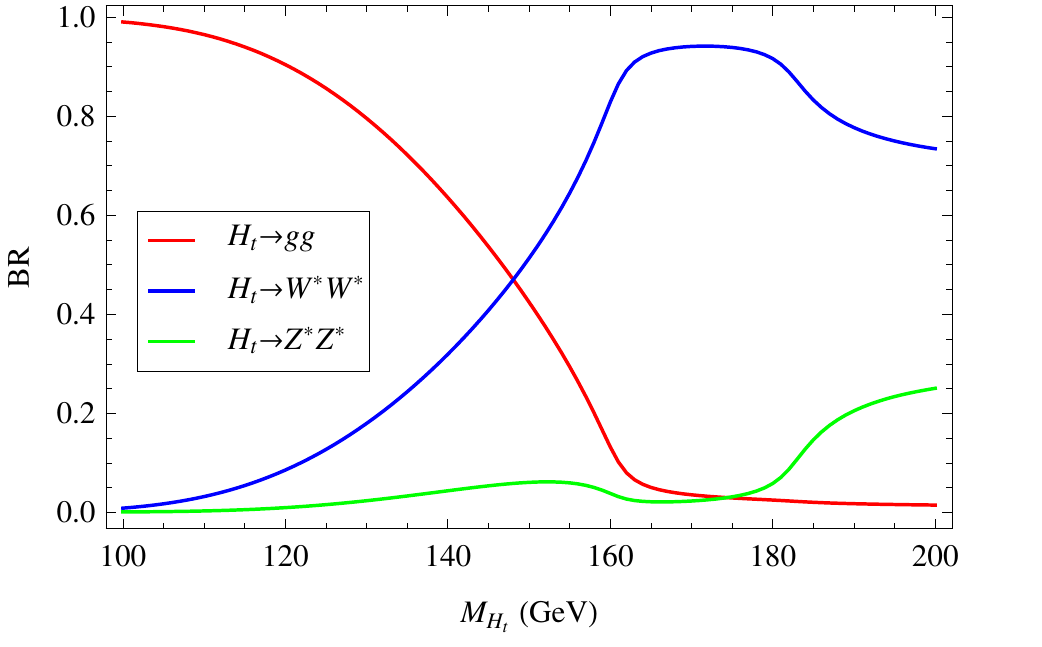}
\includegraphics[width=3.39in]{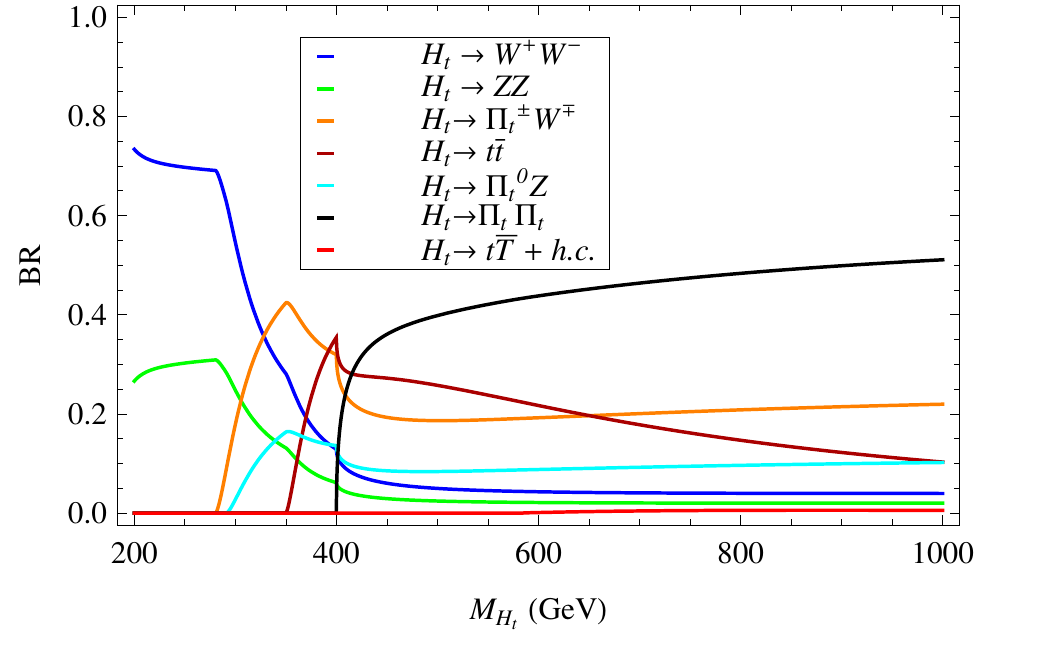}
\caption{\label{fig:br-higgs-linear}Branching ratios for the dominant decay modes of the top-Higgs $H_t$.   At left, the key shows (from top to bottom) the order of the curves at low $M_{H_t}$.   At right, the key shows (from top to bottom) the order of the curves for $M_{H_t}$ greater than the $WW$ threshold. We fixed the following: $M_D$= 400 GeV, $M_{W'}$= 500 GeV, $M_{\Pi_t}$= 200 GeV, and $\sin\omega$= 0.5.}
\end{figure}

The major two-body decay modes of the top-Higgs are the $t\bar{t}$ channel, gauge boson pair modes, and (when kinematically allowed), the $W\Pi_{t}$ and $\Pi_t\Pi_t$ modes. In Fig.~\ref{fig:br-higgs-linear}, we present a plot of the branching ratios of the top-Higgs including only the dominant decay modes for the illustrative set of parameter values:
\begin{align}
 M_{\Pi_t}&=\,200\, \textrm{GeV},\ \quad M_{D} = \,400\, \textrm{GeV} \nonumber \\ 
 M_{W'}&=\,500\, \textrm{GeV},\ \quad \textrm{sin}\,\omega =\,0.5. 
\end{align}
Note that for $M_{H_t}$ below the $WW$ threshold, the top-Higgs tends to decay to $gg$ (through a top loop), or to virtual $W$'s and $Z$'s, as shown in the left-hand pane of Fig.~\ref{fig:br-higgs-linear} (computed using a modified version of HDECAY \cite{HDECAY}).

\subsubsection{Top Higgs production: Direct}

\begin{figure}[!ht]
\includegraphics[width=4.5in]{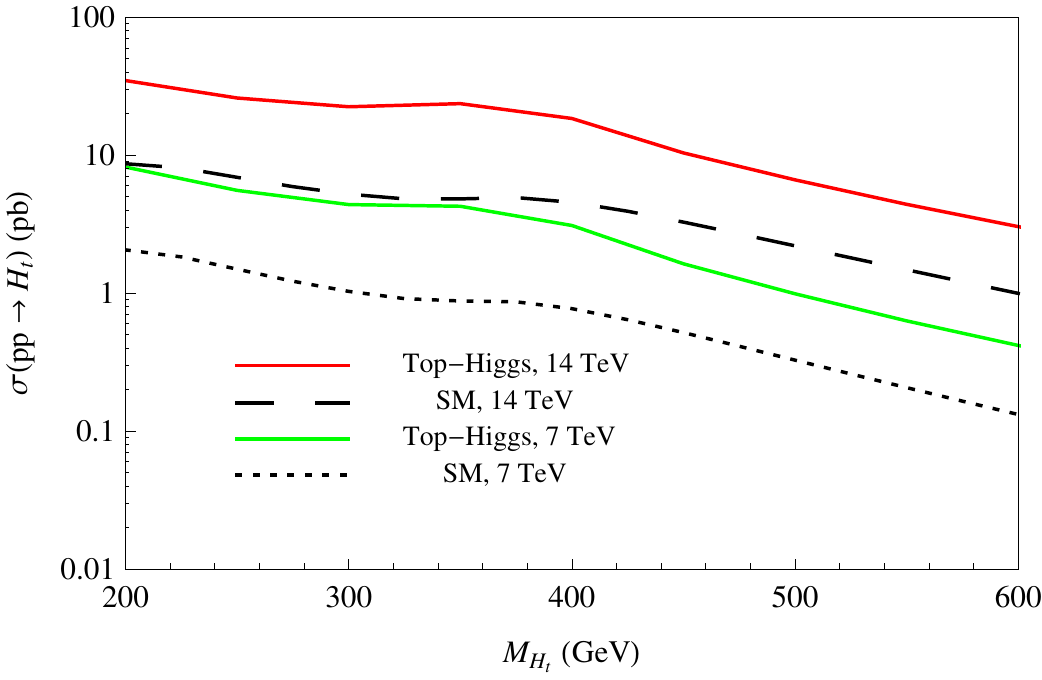}
\caption{\label{fig:ggH} The production cross-section for $pp\rightarrow H$ at the LHC. The dashed lines indicate the LO production cross sections for a standard model Higgs at  $7\ \text{TeV}$ and $14\ \text{TeV}$ at the LHC and are taken from \cite{HDECAY}, while the solid lines are the leading order (LO) top-Higgs production cross sections at the same energy values. The top-Higgs cross sections were calculated using CTEQ5L parton distribution functions, with factorization scale $\mu_F = m_{H_t}/2$ and renormalization scale $\mu_R = m_{H_t}$.}
\label{fig:ggh}
\end{figure}

The direct production of the top-Higgs, $pp \to H_t$ occurs at the LHC via gluon fusion (Fig.~\ref{fig:feynman1}, left) just as for its SM counterpart.  Within the Top-Triangle Moose model this process is completely dominated by loops of top quarks; the heavy top contribution is negligibly small. The production cross sections at the LHC are presented in Fig.~\ref{fig:ggH} for two different center of mass energies, $\sqrt s = 7\ \text{TeV}$ and $\sqrt s = 14\ \text{TeV}$.

As expected, the top-Higgs cross section is significantly larger than that for a standard model Higgs of equivalent mass. The enhancement is roughly a factor of four for our current parameter choice, though the actual value does depend somewhat on the width, and hence the mass, of the top-Higgs. Once the top-Higgs is sufficiently heavy that it can decay into a pair of top pions it becomes considerably wider than its SM equivalent, bringing down the cross section slightly.


\subsubsection{Top Higgs production: Indirect}

\begin{figure}[!ht]
\includegraphics[width=4.5in]{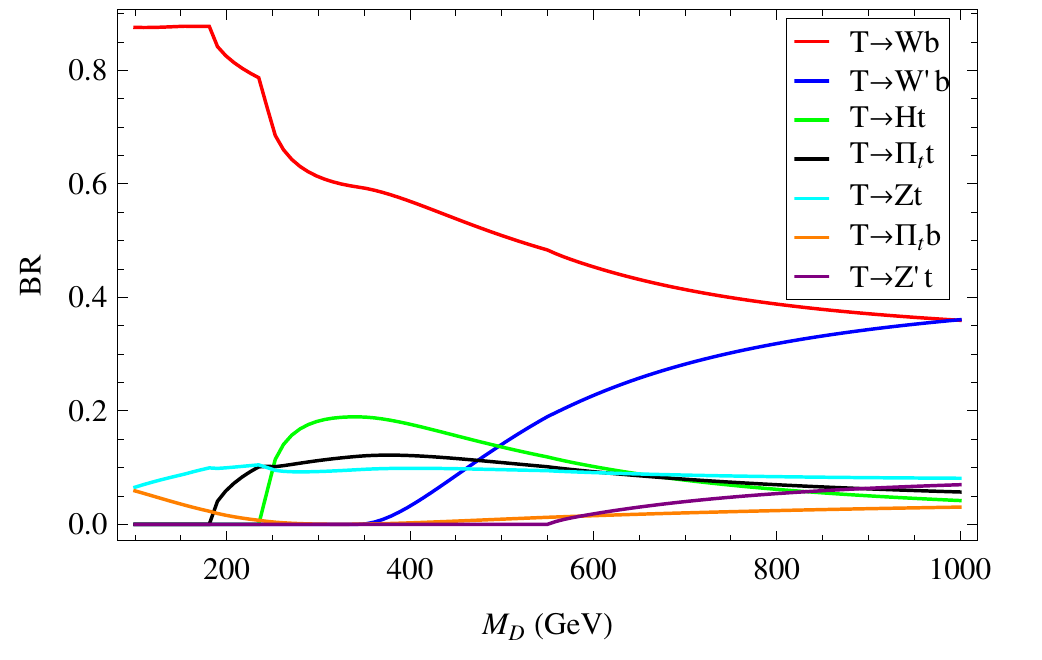}
\caption{\label{fig:br-T}The decay branching ratios of the heavy top for $M_{H_t}$= 250 GeV, $M_{\Pi_t}$= 200 GeV and $M_{W'}$= 500 GeV. The key shows (from top to bottom) the order of the curves in the middle of the plot. We see that the  decay modes involving the top higgs and top pion are comparable, while the $Wb$ mode dominates for a wide range of Dirac masses.}
\end{figure}

Since the top-Higgs has a non-zero off-diagonal coupling to a light and a heavy top, we could look for it in the decays of the heavy top. To see when this strategy might be useful, we examine the decays of the heavy top, shown in Fig.~\ref{fig:br-T}.
We see that the $Wb$ mode dominates for Dirac masses up to about a TeV. This suggests that one could look at the pair production of the heavy tops, with one of them decaying to $Wb$, and the other decaying to a top-Higgs, i.e., $pp\rightarrow T\bar{T}\rightarrow WbH_tt$, as shown in Fig.~\ref{fig:feynman1}. This strategy is identical to the indirect Higgs-production mechanism proposed previously in the context of vector-like fermion extensions of the standard model~\cite{delAguila:1989ba, delAguila:1989rq, AguilarSaavedra:2006gw,Kribs:2010ii}.  To get an idea for the size of indirect top-Higgs production in the Top-Triangle Moose model, we present the rate for $pp \rightarrow Wb\,H_tt$ at the LHC (14 TeV)  in Fig.~\ref{fig:ppHH} below.

In this plot, we have fixed $M_D\, =\, 650$ GeV,  $M_{\Pi_t}\,=\, 200$ GeV, and scanned over top-Higgs mass values from $100\ \text{GeV}$ up to\footnote{We choose 600 GeV as the upper limit because the top-Higgs becomes a broad resonance beyond this point. }  $600\ \text{GeV}$.   We will discuss the implications of top-Higgs production and decay modes in more detail in sub-section~(\ref{sec:LHCdisc}).
\begin{figure}[!h]
  \begin{center}
\includegraphics[width=4.5in]{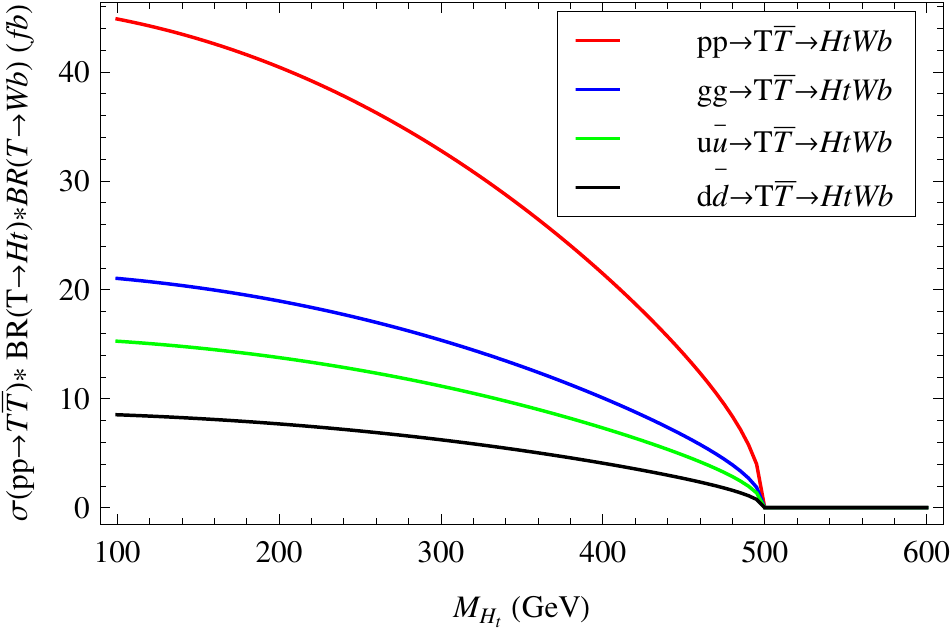}
  \end{center}
\caption{$\sigma\, \cdot\, \text{BR}$  for the process $pp\rightarrow T\bar{T}\rightarrow WbH_tt$ for $M_{\Pi_t}\, =\, 200\ \text{GeV}$, $M_{D}\, =\, 650\ \text{GeV}$, $M_{W'}\, =\, 500\ \text{GeV}$ and $\sin\omega=0.5$. This choice of final state takes advantage of the high BR for $T\rightarrow Wb$. The key shows (from top to bottom) the order of the curves in the middle of the plot. The cross section were calculated using an implementation of the Top-Triangle Moose model in CalcHEP~\cite{Pukhov-Calchep} and assumed a LHC center of mass energy of 14 TeV.}
 \label{fig:ppHH}
\end{figure}

\subsection{Top Pion production and decay}

We now turn to the top-pions. Before discussing their production channels, which are similar to the ones discussed for the top-Higgs, we will first work out the decay branching ratios of the charged and neutral top-pion.

\subsubsection{Decay Branching Ratios}

\begin{figure}[h!]
\includegraphics[width=3.4in]{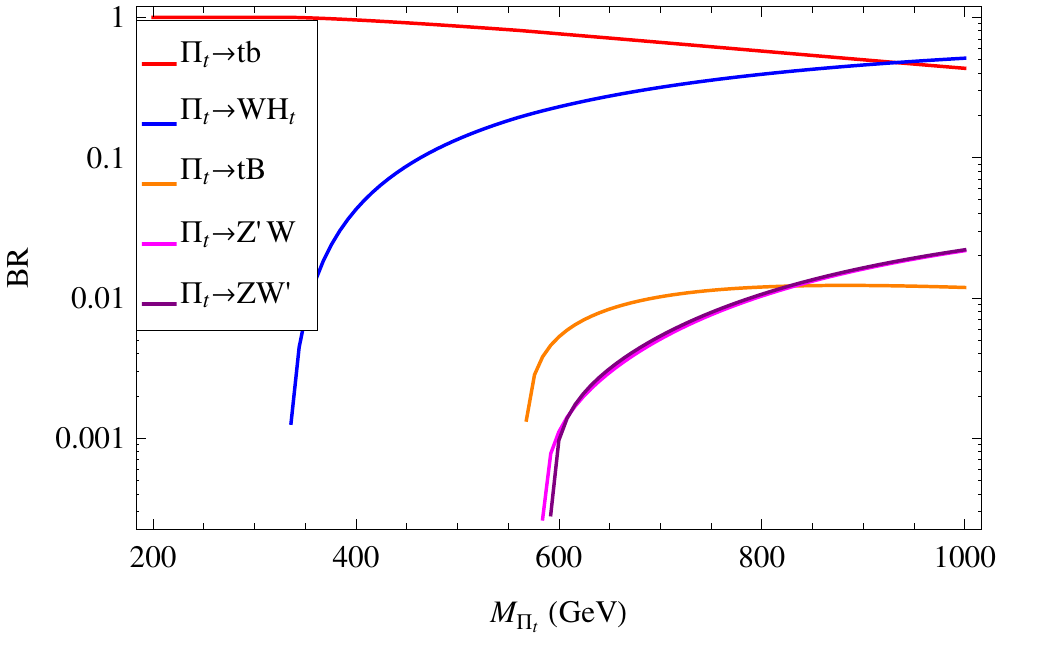}
\includegraphics[width=3.4in]{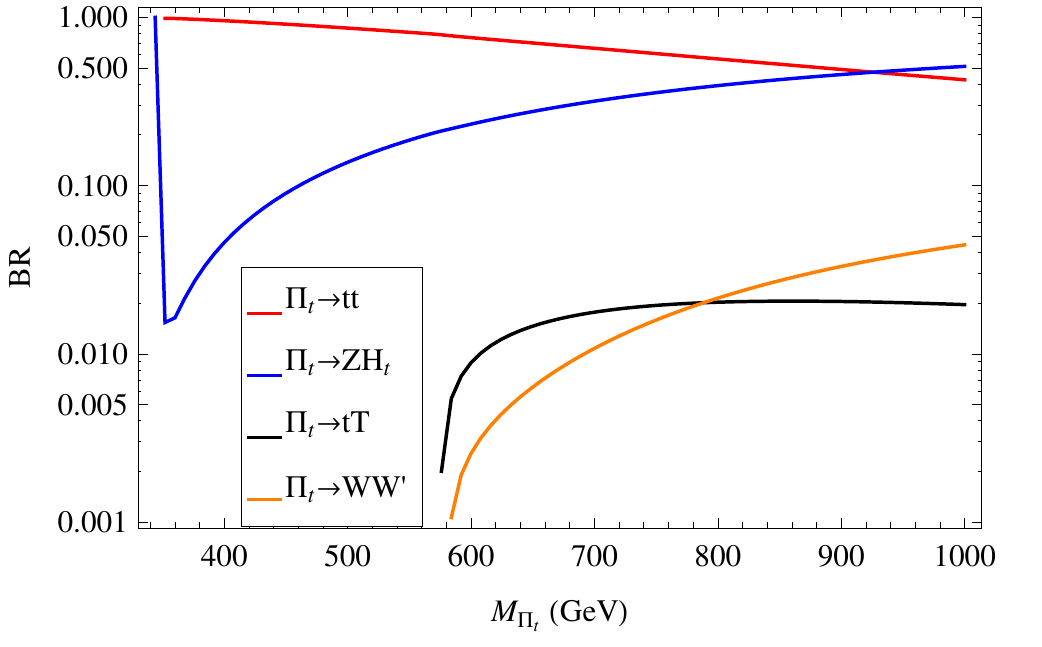}
\caption{\label{fig:br-pi}The decay branching ratios of the charged top-pions (left) and the neutral top-pion (right). The key shows (from top to bottom) the order of the curves in the middle of the plot. The dominant decay modes are to $tb, tB, WH_t, ZW', \textrm{and} WZ'$ for the $\Pi^{\pm}_t$, and $tt, tT, WW', \textrm{and} \, ZH_t$ for the $\Pi^0_t$. Below the $t \bar t $ threshold, the $\Pi^0_t$ will decay almost exclusively into a pair of gluons. }
\end{figure}

The charged (neutral) top pion, when produced, decays to $tb$ ($tt$), $Wh$ ($Zh$), or $Tb$ ($tT$). The decays involving heavy gauge bosons or two heavy fermions are suppressed. We show the plot of branching ratios of the $\Pi_{t}^{-}$ and $\Pi^0_{t}$ in Fig.~\ref{fig:br-pi} for the following illustrative set of parameter values:
\begin{align}
 M_{H_t}&=\,250\, \textrm{GeV},\ \quad M_{D} = \,400\, \textrm{GeV} \nonumber \\ 
 M_{W'}&=\,500\, \textrm{GeV},\ \quad \textrm{sin}\,\omega =\,0.5. 
\end{align}
 
As $\Pi_t$  is a pseudoscalar, it cannot decay into longitudinally polarized gauge bosons. With the longitudinal $W/Z$ modes forbidden, the dominant decay mode of $\Pi_t$ below the top-pair threshold is $\Pi_t \rightarrow gg$. Decays to pairs of (transversely polarized) electroweak bosons are present but suppressed by small coupling. Similarly, phase space suppresses three and four-body decay modes like $\Pi_t \rightarrow \bar t t^*$. As a result, the neutral top-pion is quite narrow below the top-pair threshold.

\subsubsection{Top pion: Direct, indirect  and associated single production}

\begin{figure}[h!]
\includegraphics[width=4.5in]{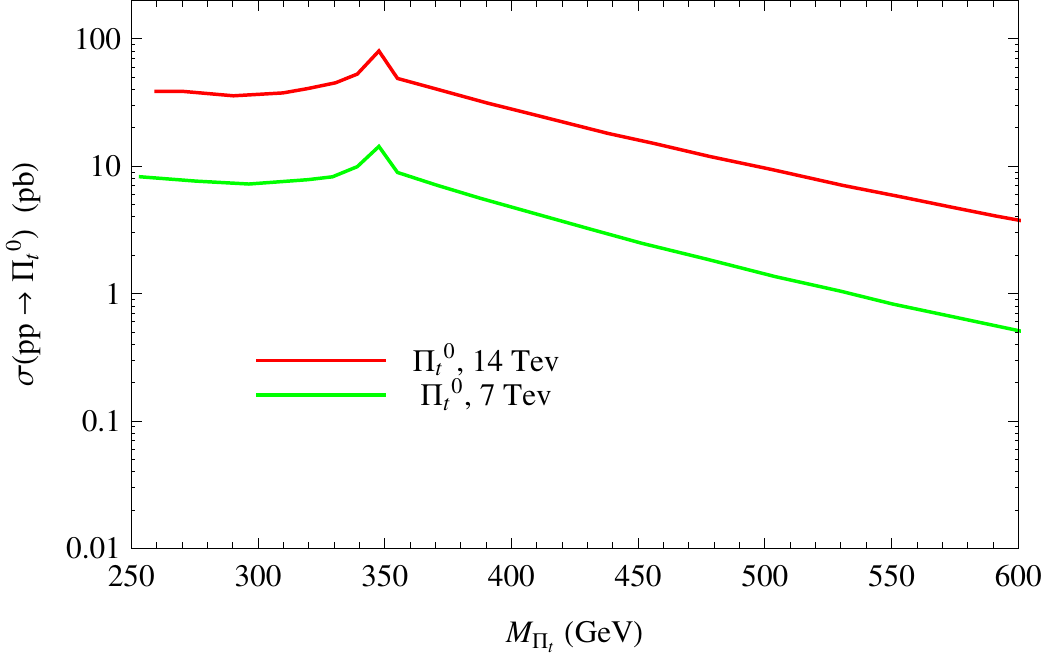}
\caption{\label{fig:ggP} The production cross-section for $pp\rightarrow \Pi_{t}$ at the LHC at 14 TeV (top curve) and 7 TeV (bottom curve). Parton distribution functions and parameters are the same as in Fig.~\ref{fig:ggh}.}
\end{figure}

\begin{figure}[htp]
  \begin{center}
  \includegraphics[width=4.5in]{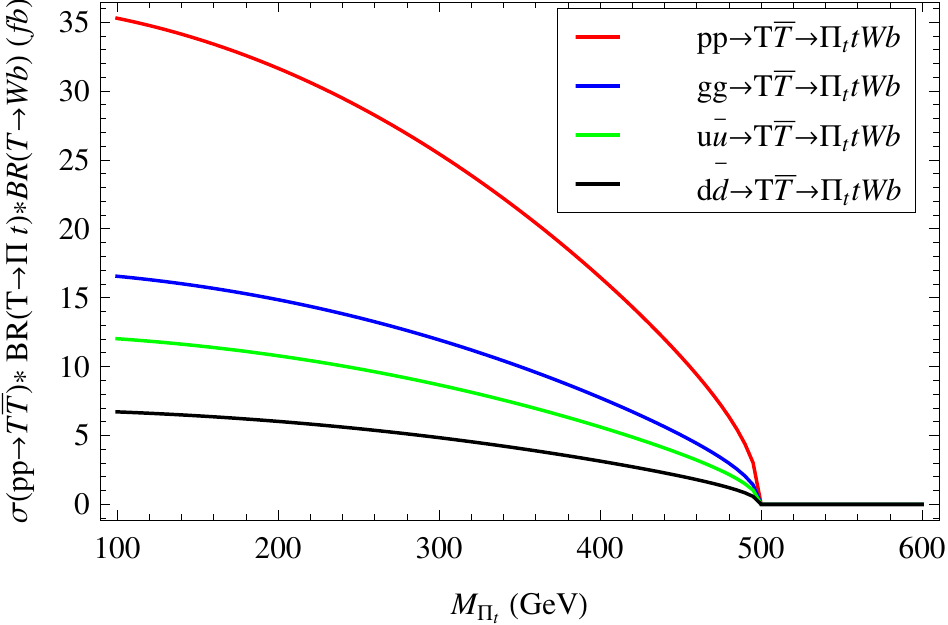}
  \end{center}
\caption{$\sigma\cdot$BR for the process $pp\rightarrow T\bar{T}\rightarrow \Pi_{t}tWb$. The  final state was chosen to take 
advantage of the high BR for $T\rightarrow Wb$. The key shows (from top to bottom) the order of the curves. This plot was made using the same tools and assumptions as Fig.~\ref{fig:ppHH}.}
 \label{fig:ppPP}
\end{figure}

The neutral top pion, by analogy with the top-Higgs, can either be produced directly via $gg\rightarrow \Pi_t$, or could show up as a decay product of the heavy top quarks. The production cross section for the first process $gg\rightarrow \Pi_{t}$ is shown in Fig.~\ref{fig:ggP} for two different LHC energies. As with top-Higgs production, the top-quark loop contribution is dominant. We see that there is a small sharp peak at $M_{\Pi}\sim 350\ \text{GeV}$ - this is due to the effect of the $t\bar{t}$ in the loop going on-shell.  In Fig.~\ref{fig:ppPP}, we present $\sigma \cdot$ BR for indirect production, again looking at the case where one of the heavy tops decays to $Wb$ and the other decays to $\Pi_{t}t$. Here, we fixed $M_D\, =\,650$ GeV, and $M_{H_t}\, =\, 250$ GeV.

\begin{figure}[]
  \begin{center}
 \includegraphics[scale=1.0]{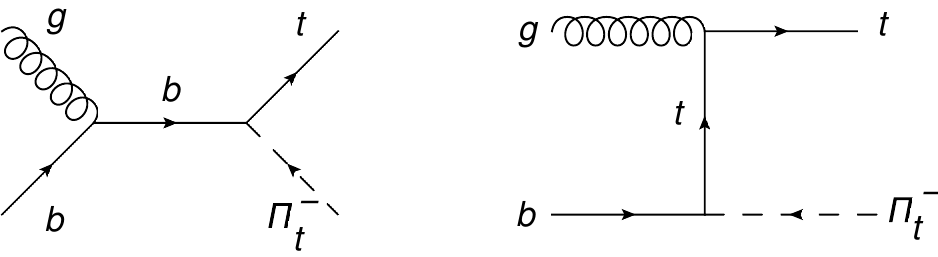}
  \end{center}
\caption{Feynman diagrams for the process $pp\to \Pi^{-}_t t$ at the LHC ($\sqrt s = 14$ TeV).}
 \label{fig:tP-Feynman}
\end{figure}

\begin{figure}[]
  \begin{center}
 \includegraphics[width=4.5in]{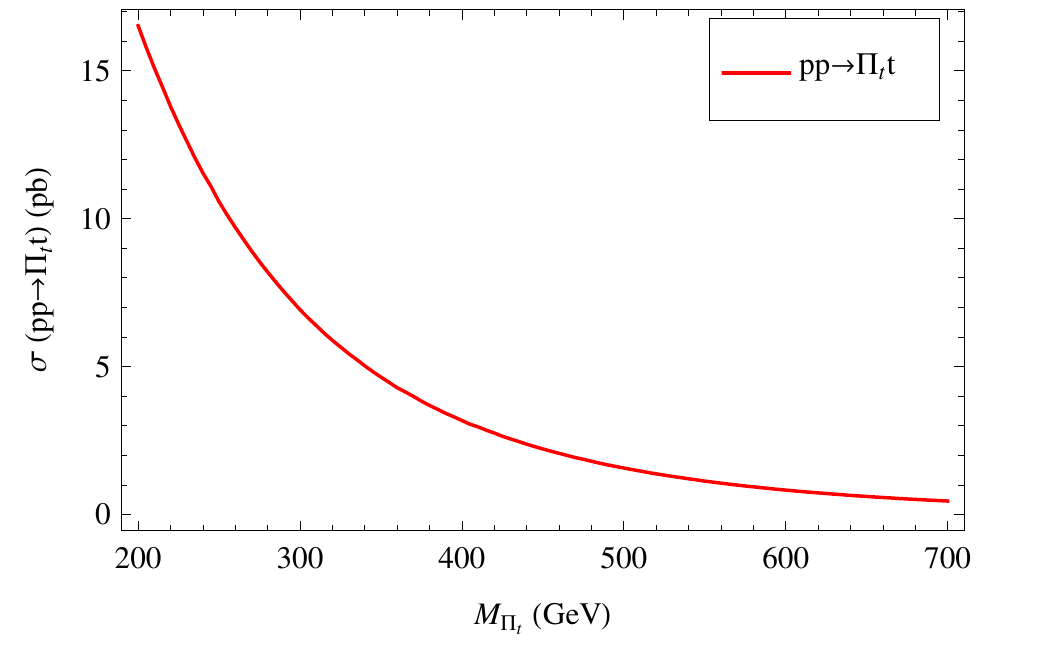}
  \end{center}
\caption{The cross-section for the process $pp\to t\Pi_t^{\pm}$ at the LHC ($\sqrt s = 14$ TeV) as a function of the top-pion mass.}
 \label{fig:tP}
\end{figure}

In addition, the top-pion can also be produced in association with a top-quark - see Fig.~\ref{fig:tP-Feynman}. We present the cross-section for this process in Fig.~\ref{fig:tP} as a function of the top-pion mass, summing over $\Pi_{t}^+$ and $\Pi_{t}^-$ production.

\subsubsection{Pair production of $H_t$ and $\Pi_t$}

\begin{figure}[h!]
\includegraphics[width=2.0in]{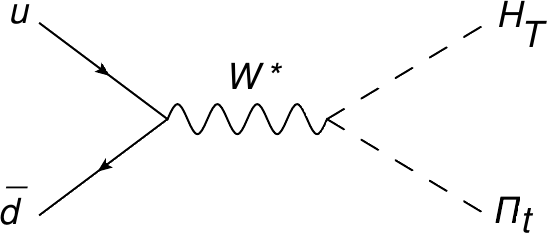}
\caption{\label{fig:higgs-pion} Feynman diagram for the production of $H_t+\Pi_{t}$} - the process occurs through an $s$ channel $W^*$.
\end{figure}

\begin{figure}[h!]
\includegraphics[width=3.5in]{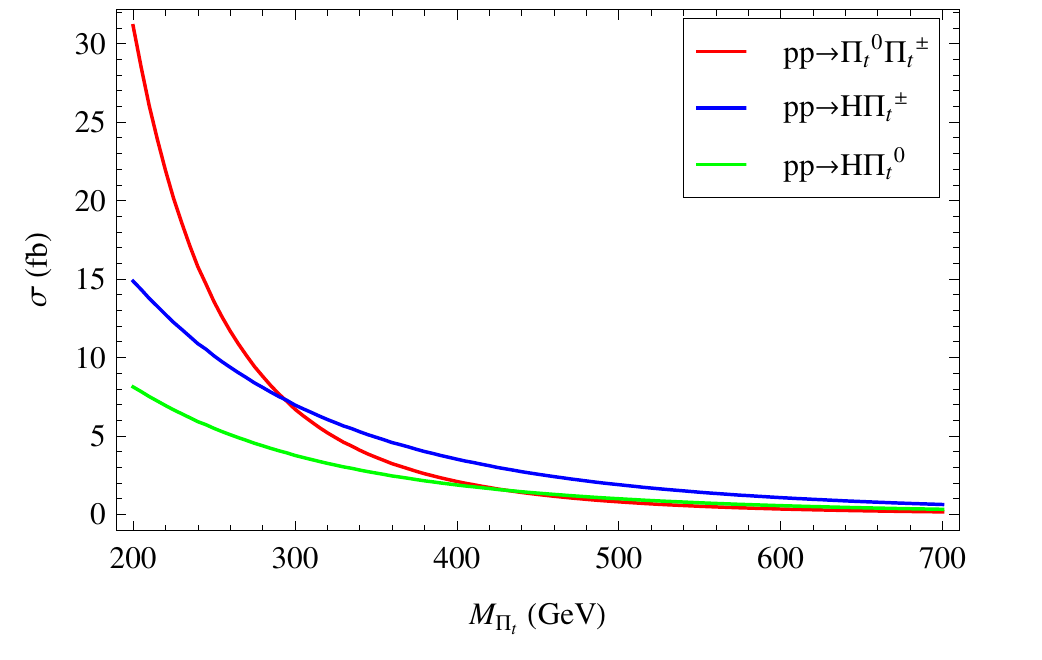}
\includegraphics[width=3.5in]{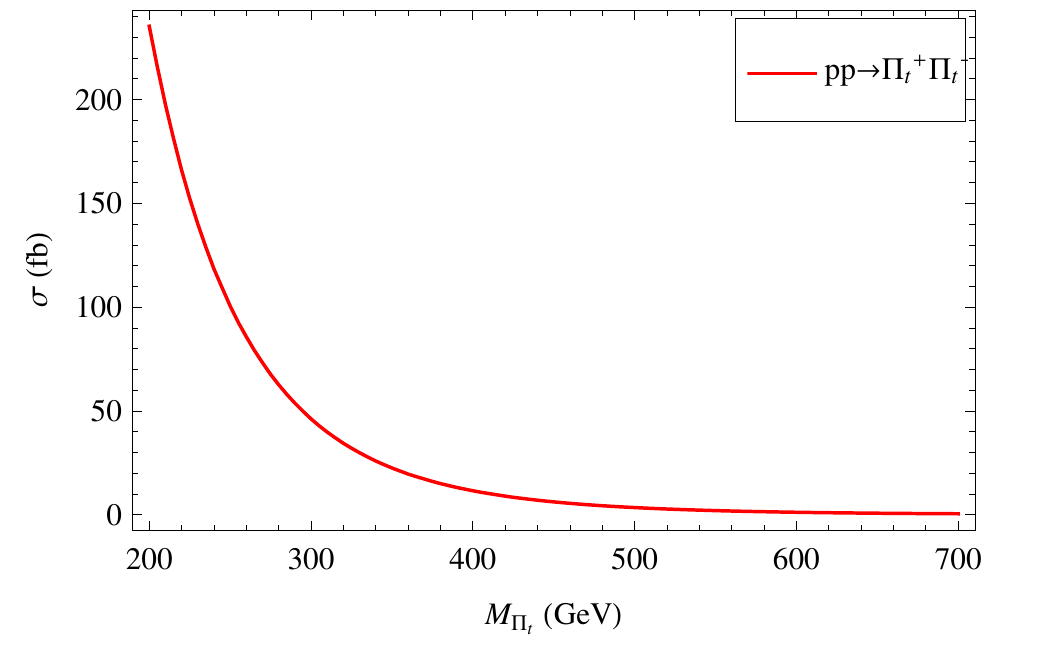}
\caption{\label{fig:pp-HP}Left:  The production cross-section for the processes $pp \to \Pi_t H_t$ and $pp \to \Pi^{0}_t\Pi^{\pm}_t$ for $M_{H_t}$= 250 GeV. The key shows (from top to bottom) the order of the curves.  Right: Cross-section for the pair productions of charged top-pions - this is seen to be signigicantly higher than the values in the plot on the left because of the contribution of additional diagrams involving the top-quark (and its heavy partner) in the $t$ channel once we include bottom quark pdf's. Both plots were made assuming a $\sqrt s = 14$ TeV LHC.}
\end{figure}

In addition to the processes considered above, one could also look at pair production of  two top-pions  or production of one top-pion and one top-Higgs at the LHC. The latter occurs via a $W^{*}$ exchange, e.g. the process $pp \rightarrow W^{*}\rightarrow \Pi_{t}^{\pm}H_t$; see Fig.~\ref{fig:higgs-pion}. We present the cross-section for this process on the left-hand side in Fig.~\ref{fig:pp-HP} as a function of the top-pion mass (keeping $M_{H_t}=$250 GeV) and summing over the $\Pi_{t}^+$ and $\Pi_{t}^-$ production. We also show the pair production of a neutral and a charged top-pion in the same plot. We have isolated the pair production of charged top-pions (the right-hand pane in Fig.~\ref{fig:pp-HP}) - one can see that the cross-section for this process is higher than the rest. This is because of the contribution of additional $t$-channel diagrams involving the top-quark (and its heavy partner) when we include the bottom quark parton distribution function.

\subsection{Discovery prospects at the LHC}
\label{sec:LHCdisc}
Now that we have discussed the production and decay of the top-Higgs, top-pion and the heavy $T$-quark in the model, we survey their discovery prospects at the LHC. We identify channels with clear discovery prospects and estimate their LHC reach. We also point out which channels are promising enough to warrant detailed investigation in future work.
Since heavy scalars can be produced indirectly, through the decay of the heavy $T$-quark, we start by commenting on the visibility of this heavy fermion at the LHC. \\


\textbf{\textit{Heavy $T$-quark}}: The LHC phenomenology of the heavy partners of the first and second generation quarks in this model was already discussed in \cite{Chivukula:2009ck} - the essential conclusion of that analysis is that, by considering both single and pair productions and subsequently letting the quarks decay to SM gauge bosons, we can discover them at the 5$\sigma$ level for a 14 TeV LHC with $\approx$ 300 $\text{fb}^{-1}$ luminosity for masses up to $\sim$ 1 TeV. Light masses, naturally, require less integrated luminosity.

Here, we discuss the prospects for discovering the heavy partner of the top-quark.  The $T$ state decays predominantly to $Wb$ for a wide range of $M_D$. Thus, for a wide range of the heavy-T mass, the best possible discovery channel, based on branching fraction considerations, seems to be $pp\to TT \to WbWb$, with the $W$'s decaying to either leptons or quarks. If both $W$'s decay leptonically, we would have two sources of missing energy, and reconstructing the heavy-$T$ mass would be problematic. Hence, the best bet\footnote{We could also consider one or both of the heavy quarks decaying to a $Z$, but this would introduce extra top decays in the final state, and is not likely to compete with the charged current channel.} seems to be $pp\to TT \to WbWb \to l\nu +4j$. But in order to facilitate comparison with  \cite{Chivukula:2009ck}, we first consider the process  $pp\to \bar TT \to WbWb \to l\nu l\nu jj$, ignoring for the moment the complication arising due to the presence of two neutrinos in the final state.  In order to make definitive statements regarding discovery prospects, we would have to calculate the complete SM background. But it is conceivable that once we impose hard $p_T$ cuts on the jets, the SM background reduces to almost zero, as was the case in \cite{Chivukula:2009ck}.  In this case, one could translate the results of that analysis by scaling the couplings. Thus, comparing the process of interest to one that was analyzed, we see that the particular ratio we are after is:
\begin{equation}
 \frac{pp\to T\bar{T} \to WbW\bar{b} \to l\nu l\nu jj}{pp\to Q\bar{Q} \to WZjj \to l\nu lljj}=\frac{BR(T\to Wb)^2}{BR(Q\to Wj) BR(Q\to Zj)}\frac{BR(W\to l\nu)}{BR(Z \to ll)}.
\end{equation}

The branching ratios of the heavy quarks depend on the Dirac mass, but we can still make rough estimates. Comparing the branching ratio plot Fig.~\ref{fig:br-T} to the one for the heavy-$U$ in \cite{Chivukula:2009ck}, we see that the branching ratio to $Wj$ is enhanced for the heavy-$T$, while that to $Zj$ is suppressed by roughly the same amount. Also, $BR(Q\to Wj)\approx 2 BR(Q\to Zj)$, as can be readily verified from Fig. 3 in \cite{Chivukula:2009ck}. These two facts mean that the first ratio in the above equation is $\geq$2. The second ratio is approximately 3.2 (using the SM values: $BR(W\to l\nu=$0.108, $BR(Z \to ll=$0.033)). Thus, we see that the reach for the heavy-$T$ is roughly enhanced by a factor of 6. But in the analysis for the heavy-$U$ quarks, there is a factor of 4 included (for the heavy partners of the first two generations), and thus in our comparison, we have to divide out by the same factor. This gives an enhancement of 1.5. Considering all this, it is conservative to estimate that the reach for the heavy-$T$ quarks, via pair production at the LHC, would be comparable to that of the heavy-$U$, and that the analysis of the pair production scenario in  \cite{Chivukula:2009ck} applies here. Thus, referring to Fig. 12 in  \cite{Chivukula:2009ck}, we conclude that, for a fixed $M_{W'}$= 500 GeV, the heavy-$T$ is discoverable at the LHC with a luminosity of 1 $\text{fb}^{-1}$ for masses up to 450 GeV. This reach is extended to about 650 (850) GeV for 10 (100) $\text{fb}^{-1}$.  This indicates that it would be worth doing a thorough analysis of the signal and background for the search for the $T$ states; we plan to present this in forthcoming work.\\

\begin{figure}[h]
\centering
\includegraphics[width=3.5in]{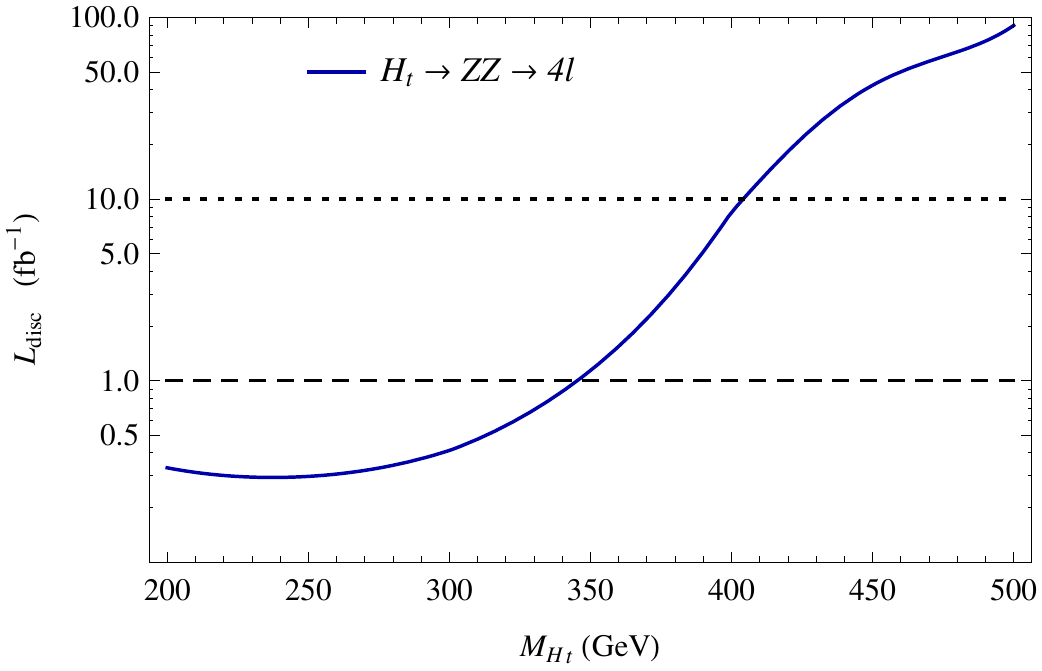}
\caption{Luminosity (in $\text{fb}^{-1}$) necessary for the discovery the top-Higgs via the fully leptonic mode $pp \rightarrow H_t \rightarrow Z\,Z \rightarrow 4\ell$ at a $14\ \text{TeV}$ LHC\@.}
\label{fig:hzzdisc}
\end{figure}

\textbf{\textit{Top-Higgs}}: Much as with the standard model Higgs, the detection prospects of the top-Higgs depend on its mass. Top-Higgses lighter than $\sim 150\ \text{GeV}$ decay dominantly into two gluons and will be impossible to see unless produced in association with a vector boson. Even when produced with a $W/Z$, the immense SM $W/Z + \text{jet}$ backgrounds would make detection difficult, especially for lighter top-Higgses\footnote{Amusingly, the CDF collaboration does see a slight excess in the di-jet invariant mass distribution of $W + \text{jets}$ events at $\sim 150\ \text{GeV}$~\cite{CDFwjj}. Though it is unlikely that the top-Higgs can be produced with sufficient rate to explain this excess, further study may be warranted.}. Above $160\ \text{GeV}$, top-Higgses produced via gluon fusion are detectable through leptonic $WW/ZZ$ modes. Gluon fusion to top-Higgses is enhanced by $1/\sin^2{\omega} \sim 4 $ over a SM Higgs of equivalent mass, making the discovery prospects excellent. To get some idea of the accessible parameter range we can rescale SM Higgs discovery projections to account for the altered production rate and decay of the top-Higgs. This is most easily done for a 14 TeV collider, where many studies have been done for all Higgs masses~(see, for example~\cite{Djouadi:2005gi}). As an example, we can concentrate on top-Higgses heavier than $200\ \text{GeV}$ where the 4-lepton `golden' mode will be dominant. The $h \rightarrow ZZ$ significances found in \cite{Cranmer:2004ys,Djouadi:2005gi} are rescaled, then translated into a luminosity required for $S/\sqrt B = 5.0$ at a given top-Higgs mass. This gives us the top-Higgs discovery luminosity curve, which we show in Fig.~(\ref{fig:hzzdisc}).
Discovery of top-Higgses ligher than $350\ \text{GeV}$ using the leptonic mode alone is possible over a wide range of masses; top-Higgses with mass $ < 350\ \text{GeV}\ (< 400\ \text{GeV})$ would be seen within the first $\text{fb}^{-1}$ ($10\ \text{fb}^{-1}$). Using the leptonic $H_t \rightarrow WW \rightarrow 2\ell\, 2\nu$ mode, we expect similar number for discovery prospects extending down to $m_{H_t} \sim 160\ \text{GeV}$.  

While the discovery prospects of a $\sim160 - 400\ \text{GeV}$ top-Higgs at a full-powered LHC are excellent, one may ask what the discovery prospects are during the initial, low-energy LHC run. Few phenomenology studies have been carried out for SM Higgs at this lower energy, however Ref.~\cite{Berger:2010nc} has studied the leptonic $WW$ mode for a 7 TeV collider for Higgses lighter than $200\ \text{GeV}$ - however, as can be seen from Fig.~\ref{fig:limit-sin.pdf}, $145\,\textrm{GeV}<M_{H_t}<195\,\textrm{GeV}$ has already been excluded by the Tevatron for $\sin\omega\leq0.5$. 
A more thorough investigation of light top-Higgses including other modes would be interesting, however from this simple rescaling alone we can say that light top-Higgses are certainly detectable even during the initial LHC run.

For heavier top-Higgses, discovery becomes more challenging because the $t \bar t$ mode opens up. In the SM the $t \bar t$ mode is never discussed as a discovery mode since the Higgs branching fraction to $t \bar t$ never gets bigger than 10\% and cannot compete with the cleaner leptonic $WW/ZZ$ channels. In contrast, the branching fraction to $t \bar t$ for the top-Higgs can be much bigger than 10\% because of the enhanced top - $H_t$ coupling and, consequently, the $WW/ZZ$ branching fractions drop at high $M_{H_t}$ much more sharply than in the SM~\cite{Djouadi:2005gi}. This, unfortunately, has a net negative effect on the discovery potential:  $H_t \rightarrow t \bar t$ is unlikely to be a discovery mode due to large backgrounds and the signal size in the cleaner di-boson channels is reduced. 

A better option for the discovery of heavy $H_t$ is  $H_t \rightarrow \Pi_t^{0} Z$. Provided that $M_{\Pi_t} < 2 m_t$, this decay mode yields the final state $\ell\ell jj$, where the jets are quite energetic and can be reconstructed to $M_{\Pi_t}$. Rejecting events with $\gtrsim 2 $ jets or heavy flavor and by exploiting the kinematics of the $\Pi_t \rightarrow jj$ system, it may be possible to suppress SM $Z + \text{jets}$ and $t \bar t$ backgrounds to the point that the top-Higgs is discoverable. While this channel can arise any time there are multiple sectors which break EWS (such as 2HDM), we are not aware of any phenomenological studies.  We plan to address this in forthcoming work.  Note that the $Z\Pi_{t}^{0}$ mode is only potentially useful for top-Higgses lighter than  $2 M_{\Pi_t}$; above $2 M_{\Pi_t}$, top-Higgses will decay primarily to $H_t \rightarrow \Pi_t\,\Pi_t$, where the top-pions can be charged or neutral. In either case, this final state will be extremely challenging to discover~\cite{Chivukula:1990di, Chivukula:1991zk, Kilic:2008ub}. \\

\textbf{\textit{Charged top-pion}}: The charged top-pion is phenomenologically similar to a charged Higgs boson in a two-Higgs-doublet model with low $\tan\beta$ (i.e., enhanced top quark Yukawa coupling).  Discovery prospects for a charged Higgs boson have previously been studied for the 14~TeV LHC in the context of supersymmetric models.  The charged top-pion can be produced in association with a top quark through bottom-gluon fusion, $gb \to t \Pi_t^-$, and through gluon-gluon fusion, $gg \to \bar b t \Pi_t^-$.  The cross section has been computed to next-to-leading order in QCD~\cite{Plehn:2002vy}; it grows proportional to $\cot^2\omega$ (analogous to $\cot^2\beta$ in the usual two-Higgs-doublet notation).  

Due to the popularity of supersymmetric models, studies of $tH^-$, $H^- \to \bar t b$ at ATLAS~\cite{Aad:2009wy} and CMS~\cite{CMSnote} have focused entirely on the large $\tan\beta$ regime.  
The major background comes from $t \bar t$ plus jets; the systematic uncertainty from the background normalization presents the biggest challenge to this search. 
The CMS study~\cite{CMSnote} gives values of $\sigma(pp \to t H^{\pm}) \times {\rm BR}(H^{\pm} \to tb)$ required for $5\sigma$ discovery in this channel as a function of the mass of the pseudoscalar $A^0$ in the MSSM, which is nearly degenerate with $H^-$ in that model - the sensitivity depends very strongly on the systematic uncertainty on the $t\bar{t}$ background. We present a plot of the cross-section for the process $pp\to t\Pi^{-}_t$ in Fig.~\ref{fig:tP}. For $M_{\Pi_t}\leq600\,\textrm{GeV}$, the charged top-pion decays to $tb$ 100\% of the time (see Fig.~\ref{fig:br-pi}), and so we can make a direct comparison between Fig.~\ref{fig:tP} and the CMS study. Doing so, we find that one can discover a charged top-pion at the 5$\sigma$ level at a luminosity of 30 $\text{fb}^{-1}$ for $M_{\Pi_t}\leq450\,\textrm{GeV}$, assuming a 0\% uncertainty on the $t\bar{t}$ background\footnote{Note that $\sigma(pp\to\Pi^{-}_t t)\propto \cot^2\omega$, and hence the reach becomes higher for lower $\sin\omega$.}. For 1\% (3\%) systematic uncertainty, the reach goes down to 350 GeV (250 GeV)\footnote{The CMS study only looks at charged Higgs of mass at least 250 GeV.}.

Even when decays to other final states ($WH_{t}$, $tB$, $ZW^{\prime}$, $Z^{\prime} W$) are kinematically accessible, the branching fraction to $\bar t b$ remains high - see Fig.~\ref{fig:br-pi}.  Studies of this channel done in the context of the MSSM can thus still be applied, with the caveat that angular correlations among the final-state particles in the event may be different.  In the MSSM at large $\tan\beta$, $H^-$ decays to $\bar t_L b_R$ through the bottom Yukawa coupling.  For the top-pion, however, $\Pi_t^- \to \bar t_R b_L$ through the top Yukawa coupling.  This difference may affect details of the experimental acceptance for the signal.  However, it is reasonable to conclude that this channel is quite promising for $\Pi_t$ discovery and warrants further study.

Finally we note that ATLAS~\cite{Aad:2009wy} combines this $tH^-$, $H^- \to \bar t b$ channel with $tH^-$, $H^- \to \tau \nu$ in the MSSM to present combined discovery reach contours at large $\tan\beta$, but the $H^- \to \bar t b$ contribution improves the reach only marginally~\cite{Aad:2009wy}. We emphasize that while BR($H^- \to \tau \nu) \simeq$ 10\% above the $tb$ threshold in the MSSM at large $\tan\beta$, for the charged top-pion this decay mode is absent.

\section{Discussion and conclusions}
\label{sec:conclusions}

This paper has explored the collider physics of the new heavy fermionic top-quark partner ($T$), the top-Higgs boson ($H_t$) and the top-pion ($\Pi_t$) states in the deconstructed topcolor-assisted technicolor theory known as the Top Triangle Moose.  After establishing the spectrum and the couplings of the new states to each other and to standard model particles in Section III, we turned to phenomenology.  We showed in Section IV.A how existing Fermilab Tevatron data constrains $M_{H_t}$ and $M_{\Pi_t}$ as a function of the mixing angle $\sin\omega$ between the linear and nonlinear sigma model symmetry-breaking sectors: $\Pi_t$ lighter than 150 GeV would likely have been seen already in $t \to \Pi_t b$ while $H_t$ in the 150-200 GeV range would likely have been visible in $WW$ decays.  

We also established that the presence of relative light $\Pi_t$ does not present insurmountable challenges related to third-generation flavor physics, as one might have feared.  In particular, allowing the delocalization of the left-handed top quark to deviate from the value suggested by ideal delocalization can cancel contributions from one-loop diagrams involving $\Pi_t$ exchange that would otherwise have shifted $R_b$ from agreement with experiment.  As shown in Figure 3, nearly the full $\sin\omega$ vs. $M_{\Pi_t}$ parameter space can be accommodated in this way.  Moreover, limits from third-generation FCNCs are consistent with this finding, as shown in Appendix D.

In Sections IV.B and IV.C, we laid the foundation for studies of LHC phenomenology by calculating the decay branching ratios of $H_t$ and $\Pi_t$, as well as their production cross-sections, for a variety of key processes.  This information allowed us to determine which channels are most promising for discovery of $T$, $H_t$ and $\Pi_t$. Adapting previous work on the heavy partners $Q$ of the first and second generation quarks enabled us to demonstrate that the $T$ should be visible at the LHC for $M_T \leq 900$ GeV in the pair-production channel $pp \to T \bar{T} \to Wb Wb \to \ell\nu\ \ell\nu\ jj$.   Hence, a full study of the detailed background processes and optimal cuts for this process is indicated (and is now underway).  The alternative channel in which one $W$ decays hadronically, so that the final state is $\ell\nu\ 4j$, should offer a larger signal along with the welcome possibility for full reconstruction of the top quark's heavy partner; we are also planning to study this channel. 

In the case of a moderately light $H_t$, we found that the situation resembles that of the standard model Higgs.  For $M_{H_t} \leq 160$ GeV, the top-Higgs will be invisible because it decays almost exclusively to dijets, for which the background is overwhelming.  For $160\ {\rm GeV} \leq M_{H_t} \leq 400$ GeV, the top-Higgs should actually be easier to find than the standard model Higgs, because the ``golden" all-leptonic decay modes open up and the signal rate is enhanced by a factor of $1 / \sin\omega$.  In fact, for $H_t$ in the lighter end of this mass range, discovery in the first $1 \text{fb}^{-1}$ of LHC data would be possible.  Top-Higgs bosons heavier than 400 GeV will be more challenging to find since $BR(H_t \to WW)$ will be below even the already-reduced diboson branching ratio for the standard model Higgs.  The most promising decay channel for top-Higgs discovery in the window $2 m_t \leq M_{H_t} \leq 2 M_{\Pi_t}$ would be $H_t \to W \Pi_t$ and we plan to study this in detail in forthcoming work.  Once $M_{H_t} > 2 M_{\Pi_t}$, the primary decay mode is $H_t \to \Pi_t \Pi_t \to 4 g$ and the large multijet background will make discovery difficult (though the methods advocated in \cite{Chivukula:1990di,Chivukula:1991zk,Kilic:2008ub} can be of help).

Adapting existing work on charged-Higgs search protocols suggests that  $\Pi_t^\pm$ with masses below 400 GeV should be visible in 30 $\text{fb}^{-1}$ of LHC data through the process $pp \to t \Pi_t^\pm \to ttb$.  Further studies of final state particle angular correlations and the dependence on $\sin\omega$ are needed.  In particular, most studies of charged-Higgs searches have focused on the case of large mixing angle (essentially, large $\tan\beta$) whereas the case of small $\sin\omega$ is of greatest interest in the Top Triangle Moose.

Finally, it is interesting to reflect on how one would know that the new states one had discovered, whether $T$, $H_t$ or $\Pi_t$, were those of the Top Triangle Moose rather than some other model.  The answer will surely lie in the overall pattern of observable relationships among these three states.  Consider, for instance, a top-Higgs boson of moderate mass.  One would first find this state in single production followed by diboson decays, $pp \to H_t \to WW$; the fact that the signal rate noticeably exceeded the standard model prediction would show that one had found an exotic rather than a standard model Higgs state.  As the LHC integrated luminosity grew, the $T$ state would eventually be found in $pp \to T\bar{T} \to WbWb$ channels.  Once the existence of that state is confirmed, it would be possible to measure the rarer $T \to H_t t$ decay path and confirm that the $H_t$  found in $T$ decays is the same particle that one had already discovered in $H_t \to WW$.  This would show that the $H_t$ was both part of the electroweak sector (as witnessed by its diboson coupling) and strongly coupled to the top quark sector.  In the case of top-pions, one might begin by establishing their presence in associated production with top quarks; this would help show that they were strongly coupled to the top sector, which a measurement of $T \to \Pi_t t$ could also confirm.  Then finding either joint production of $H_t$ and $\Pi_t$ through an off-shell $W$ boson ($pp \to W^* \to H_t \Pi_t$) or one of the decay paths $H_t \to \Pi_t W$ or $\Pi_t \to H_t W$ would demonstrate the relationship of $\Pi_t$ to the electroweak sector, including the top-Higgs.

As the LHC data set grows, it will be interesting to watch for signs of these new states, heralding the presence of new strong dynamics in the top quark sector.

\begin{acknowledgments}
BC and HEL were supported, in part, by the Natural Sciences and Engineering Research Council of Canada.  RSC and EHS were supported, in part, by the US National Science Foundation under grant PHY-0854889. AM is supported by Fermilab operated by Fermi Research Alliance, LLC under contract number DE-AC02-07CH11359 with the US Department of Energy.

\end{acknowledgments}


\appendix

\section{Masses and Eigenstates}
\subsection{Gauge Bosons}

The neutral gauge boson mass matrix is given by:
\begin{eqnarray}
M_{Z}^{2}=\frac{e^{2}\, v^{2}}{4\, x^{2}\,\textrm{sin}^{2}\,\theta}\left(\begin{array}{ccc}
\frac{x^{2}}{1-x^{2}}(1+\textrm{cos}^{2}\,\omega) & -\frac{2 x}{\sqrt{1-x^{2}}}\textrm{cos}^{2}\,\omega & -\frac{x^{2}}{\sqrt{1-x^{2}}}\textrm{sin}^{2}\omega\,\textrm{tan}\,\theta \\
-\frac{2 x}{\sqrt{1-x^{2}}}\textrm{cos}^{2}\,\omega & 4\,\textrm{cos}^{2}\omega & -2\, x\textrm{\, cos}^{2}\,\omega\textrm{\, tan}\,\theta\\
-\frac{x^{2}}{\sqrt{1-x^{2}}}\textrm{sin}^{2}\omega\,\textrm{tan}\,\theta & -2\, x\,\textrm{cos}^{2}\,\omega\,\textrm{tan}\,\theta & x^{2}(1+\textrm{cos}^{2}\,\omega)\textrm{tan}^{2}\,\theta\end{array}\right).
\label{eqn:Z mass matrix}
\end{eqnarray}
Diagonalizing perturbatively in the small parameter $x$ yields the
 following masses for the $Z$ and the $Z'$~\cite{Chivukula:2009ck}:

\begin{align}
M_{Z}^{2}&=\frac{e^{2}\, v^{2}}{4\,\textrm{sin}^{2}\,\theta\,\textrm{ cos}^{2}\,\theta}\left(1+x^{2}\left(1-\frac{\textrm{sec}^{2}\,\theta}{4}\right)\right)
\label{Z mass} \\
M_{Z'}^{2}&=\frac{e^{2}\, v^{2}\,\textrm{cos}^{2}\,\omega}{4\,\textrm{sin}^{2}\,\theta\, x^{2}}\left(4+x^{2}\textrm{sec}^{2}\,\theta\right),
\label{Z' mass}
\end{align}
while the photon remains massless. The eigenvector of the $Z$ is given by:

\begin{equation}
Z^{\mu} =  v_{z}^{0}W_{0}^{\mu}+v_{z}^{1}W_{1}^{\mu}+v_{z}^{2}B^{\mu},
\label{Z eigenvector} 
\end{equation}
where
\begin{equation}
v_{z}^{0}  =  \frac{1}{8}\,(4(-2+x^{2})\textrm{cos\,}\theta-3\, x^{2}\textrm{sec}\,\theta), \,\,\,\,
v_{z}^{1}  =  \frac{1}{2}\, x(-2\textrm{cos}^{2}\,\theta+1)\textrm{sec}\,\theta, \,\, \,\,\,\,
v_{z}^{2}  =  \textrm{sin}\,\theta-\frac{1}{2}x^{2}\,\textrm{sec}\,\theta\,\textrm{tan}\,\theta. \nonumber
\end{equation}
The eigenvector of $Z'$ is the orthogonal combination. The charged gauge boson mass matrix is the upper $2\times 2$ block of Eq.~\eqref{eqn:Z mass matrix}. The masses of the physical gauge bosons are given by:

\begin{align}
 M_{W}^{2}&=\frac{e^{2}v^{2}}{4\,\textrm{sin}^{2}\,\theta}\left(1+\frac{3x^{2}}{4}\right) \\
 M_{W'}^{2}&=\frac{e^{2}\, v^{2}\textrm{cos}^{2}\,\omega}{4\,\textrm{sin}^{2}\,\theta\, x^{2}}\left(4+x^{2}\right),
\end{align}
with the respective eigenvectors:
\begin{align}
 W^{\mu} & =\left(1-\frac{x^{2}}{8}\right)W_{0}^{\mu}+\frac{1}{2}xW_{1}^{\mu}\ \textrm{and} \\
 W'^{\mu} & =-\frac{1}{2}xW_{0}^{\mu}+\left(1-\frac{x^{2}}{8}\right)W_{1}^{\mu}. 
\end{align}

We are now in a position to define the weak mixing angle, 
$1 - \sin^2\theta_W \equiv M_W^2/M_Z^2$.  Including corrections
up to $\mathcal{O}(x^2)$, we obtain,
\begin{equation}
  \sin\theta_W = \left( 1 - \frac{x^2}{8} \right) \sin\theta.
\end{equation}

\subsection{Fermions}

The light fermion mass matrix is derived from the Lagrangian:
\begin{eqnarray}
\mathcal{L} & = & M_{D}\left[\epsilon_{L}\bar{\psi}_{L0}\Sigma_{01}\psi_{R1}+\bar{\psi}_{R1}\psi_{L1}+\bar{\psi}_{L1}\Sigma_{12}\left(\begin{array}{cc}
\epsilon_{uR} & 0\\
0 & \epsilon_{dR}\end{array}\right)\left(\begin{array}{c}
u_{R2}\\
d_{R2}\end{array}\right)\right],
\end{eqnarray}
and is given by:
\begin{eqnarray}
M_{u,d}=M_{D}\left(\begin{array}{cc}
\epsilon_{L} & 0\\
1 & \epsilon_{uR,dR}\end{array}\right).
\label{fermion mass matrix}
\end{eqnarray}

This can be diagonalized in the small parameters $\epsilon_L$ and $\epsilon_{fR}$ to yield the masses of the light fermion and its heavy Dirac partner:

\begin{align}
m_{f}&=\frac{M_{D}\epsilon_{L}\epsilon_{fR}}{\sqrt{1+\epsilon_{fR}^{2}}}\left[1-\frac{\epsilon_{L}^{2}}{2(1+\epsilon_{fR}^{2})}+...\right] \label{eqn:light quark mass}\\
m_{F}&=M_{D}\sqrt{1+\epsilon_{fR}^{2}}\left[1+\frac{\epsilon_{L}^{2}}{2(1+\epsilon_{fR}^{2})^{2}}+....\right].
\end{align}

The left- and right-handed eigenstates of the light fermion can be derived to be:

\begin{align}
u_{L} & = \left(-1+\frac{\epsilon_{L}^{2}}{2(1+\epsilon_{uR}^{2})^{2}}\right)\psi_{L0}+\left(\frac{\epsilon_{L}}{1+\epsilon_{uR}^{2}}\right)\psi_{L1}, \\
u_{R} & = \left(-\frac{\epsilon_{uR}}{\sqrt{1+\epsilon_{uR}^{2}}}+\frac{\epsilon_{L}^{2}\epsilon_{uR}}{(1+\epsilon_{uR}^{2})^{5/2}}\right)\psi_{R1}+\left(\frac{1}{\sqrt{1+\epsilon_{uR}^{2}}}+\frac{\epsilon_{L}^{2}\epsilon_{uR}^{2}}{(1+\epsilon_{uR}^{2})^{5/2}}\right)u_{R2}.
\label{uR vector}
\end{align}
The eigenvector of the left- and right-handed heavy quark are the orthogonal combinations.

For the top, the mass term is dominated by the top-Higgs contribution. The mass matrix is given by:

\begin{eqnarray}
M_{t}=M_{D}\left(\begin{array}{cc}
\epsilon_{tL} & a\\
1 & \epsilon_{tR}
\end{array}\right),
\label{eqn:top mass matrix}
\end{eqnarray}
where the parameter $a$ is defined as $a \equiv v \sin\omega/\sqrt{2}\,M_D$. Diagonalizing Eq.~\eqref{eqn:top mass matrix} perturbatively in $\epsilon_{tL}$ and $\epsilon_{tR}$, we get the mass of the SM top-quark:

\begin{equation}
m_{t}= \lambda_{t}v\,\textrm{sin}\,\omega\left[1+\frac{\epsilon_{tL}^{2}+\epsilon_{tR}^{2}+\frac{2}{a}\epsilon_{tL}\epsilon_{tR}}{2(-1+a^{2})}\right].
\label{eqn:top mass}
\end{equation}
Thus, we see that $m_{t}$ depends only slightly on $\epsilon_{tR}$, in contrast to the light fermion mass, Eq.~\eqref{eqn:light quark mass}, where the dominant term is $\epsilon_{fR}$ dependent. The mass of the heavy partner of the top is given by:

\begin{equation}
m_{T}=  M_{D}\left[1-\frac{\epsilon_{tL}^{2}+\epsilon_{tR}^{2}+2a\epsilon_{tL}\epsilon_{tR}}{2(-1+a^{2})}\right].
\label{eqn:mass of TOP}
\end{equation}

The left- and right-handed eigenvectors of the SM top are given by:
\begin{align}
t_{L} & = \left(1-\frac{\epsilon_{tL}^{2}+a^{2}\epsilon_{tR}^{2}+2a\epsilon_{tL}\epsilon_{tR}}{2(-1+a^{2})^{2}}\right)\psi_{L0}^{t}+\left(\frac{\epsilon_{tL}+a\epsilon_{tR}}{-1+a^{2}}\right)\psi_{L1}^{t}\  \textrm{and}\\
t_{R} & =\left(1-\frac{a^{2}\epsilon_{tL}^{2}+\epsilon_{tR}^{2}+2a\epsilon_{tL}\epsilon_{tR}}{2(-1+a^{2})^{2}}\right)\psi_{R1}^{t}+\left(\frac{a\epsilon_{tL}+\epsilon_{tR}}{-1+a^{2}}\right)t_{R2}.
\end{align}


\section{The Lagrangian}
In order to derive the terms in the Lagrangian describing the interaction of the top-Higgs and the top pions with the gauge bosons, we start by plugging Eq.~\eqref{eqn:sigma01} in Eq.~\eqref{eqn:covariant}, and writing the covariant derivative of $\Sigma_{01}$ as
\begin{equation}
 D_{\mu}\Sigma_{01}=\frac{i}{F}\partial_{\mu}\pi_0 +igW_{0\mu}-\frac{g}{F}W_{0\mu}\pi_{0}-i\tilde{g}W_{1\mu}-\frac{\tilde{g}}{F}\pi_{0} W_{1\mu},
\end{equation}
where we have denoted $\pi_{0}=\pi_{0}^{a}\sigma^{a}$. The product can be evaluated to be:
\begin{align}
(D_{\mu}\Sigma_{01})^{\dagger}(D_{\mu}\Sigma_{01} )&=\frac{1}{F^2}\left(\partial_{\mu}\pi_{0}\right)^2+\left[g^2 W_{0\mu}^{2}+\tilde{g}^2 W_{1\mu}^2-g\tilde{g}\, W_{0}^{\mu}W_{1\mu}-g\tilde{g}\, W_{1}^{\mu}W_{0\mu}\right] \nonumber\\
&+\frac{1}{F}\left(\partial^{\mu}\pi_{0}\right)\left[gW_{0\mu}-\tilde{g}W_{1\mu}\right]+\left[\frac{g}{F}W^{0\mu}\left(\partial_{\mu}\pi_{0}\right)-\frac{\tilde{g}}{F}W^{1\mu}\left(\partial_{\mu}\pi_{0}\right)\right] \nonumber \\
&-\frac{i}{F^2}\left(\partial^{\mu}\pi_{0}\right)\left[-gW_{0\mu}\pi_{0}+\tilde{g}\pi_{0}W_{1\mu}\right]-\left[\frac{ig}{F^2}\pi_{0}W^{\mu}_{0} \left(\partial_{\mu}\pi_{0}\right)-\frac{i\tilde{g}}{F^2}W^{\mu}_{1}\pi_{0} \left(\partial_{\mu}\pi_{0}\right)\right] \nonumber \\
&+\frac{i}{F}\left[-g W^{\mu}_{0}+\tilde{g}W^{\mu}_{1}\right]\left[-g W_{0\mu}\pi_{0}+\tilde{g}\pi_{0}W_{1\mu}\right]-\frac{1}{F}\left[g\pi_{0}W^{\mu}_{0}-\tilde{g}W^{\mu}_{1}\pi_{0}\right]\left[igW_{0\mu}-i\tilde{g}W_{1\mu}\right] \nonumber \\
&+\frac{1}{F^2}\left[-g\pi_{0}W^{\mu}_{0}+\tilde{g}W^{\mu}_{1}\pi_{0}\right] \left[-gW_{0\mu}\pi_{0}+\tilde{g}\pi_{0}W_{1\mu}\right].
\label{eqn:sigma0}
\end{align}
The first line gives the kinetic energy term for the pions and the gauge bosons masses. The second line gives the mixing between the gauge and the Goldstone bosons. The third and fourth lines give the $\pi\pi V$ and $\pi V V$ couplings respectively, while the last line gives the four point coupling, $\pi \pi V V$. Plugging in the matrix definitions of the fields and taking the trace, we get,

\begin{equation*}
	\addtolength{\fboxsep}{10pt}
			\boxed{
				\begin{split}
					&\mathcal{L}^{(2)}_{\pi KE}=\frac{1}{2}\left(\partial_{\mu}\pi_{1}^{a}\right)^2\\	
                                        &\mathcal{L}^{(2)}_{\textrm{mixing}}=\frac{F}{2}\left[\tilde{g}W_{1\mu}^{a}(\partial_{\mu}\pi_{1}^{a})-g'B_{2\mu}(\partial_{\mu}\pi_{1}^{3})\right]\\
                                        &\mathcal{L}^{(2)}_{\pi \pi V}=-\frac{\tilde{g}}{2}\epsilon_{abc}(\partial^{\mu}\pi_{1}^{a})W_{1\mu}^{b}\pi_{1}^{c}+\frac{g'}{2}\epsilon_{ab3}(\partial^{\mu}\pi_{1}^{a})\pi_{1}^{b}B_{2\mu}\\
                                        &\mathcal{L}^{(2)}_{\pi VV}=\frac{g'\tilde{g}F}{2}\epsilon_{ab3}W_{1}^{a\mu}B_{2 \mu}\pi_{1}^{b}\\
                                        &\mathcal{L}^{(2)}_{\pi \pi VV}=\mathcal{M}_{abcd}\tilde{g}^{2}\pi_{1}^{a}W_{1}^{b\mu}W_{1\mu}^{c}\pi_{1}^{d}-2\tilde{g}g'\mathcal{M}_{abc3}\pi_{1}^{a}W_{1}^{b\mu}\pi_{1}^{c}B_{2\mu}+\frac{g'^{2}}{8}B_{2\mu}^2(\pi_{1}^{a})^{2}\\
				\end{split}
			}
	\end{equation*}
where $\mathcal{M}_{abcd}=\frac{1}{8}\left(\delta_{ab}\delta_{cd}-\delta_{ac}\delta_{bd}+\delta_{ad}\delta_{bc}\right)$.

The corresponding terms from the kinetic term of the other nonlinear sigma model field can be read off by relabeling the fields and couplings as follows:

\begin{equation}
gW_{0\mu}\rightarrow \tilde{g}W_{1\mu};\, \tilde{g}W_{1\mu}\rightarrow g'B_{2\mu};\, \pi_{0}\rightarrow \pi_{1}.
\end{equation}

We summarize the results for the sake of completeness:

\begin{equation*}
	\addtolength{\fboxsep}{10pt}
			\boxed{
				\begin{split}
					&\mathcal{L}^{(2)}_{\pi KE}=\frac{1}{2}\left(\partial_{\mu}\pi_{1}^{a}\right)^2\\	
                                        &\mathcal{L}^{(2)}_{\textrm{mixing}}=\frac{F}{2}\left[\tilde{g}W_{1\mu}^{a}(\partial_{\mu}\pi_{1}^{a})-g'B_{2\mu}(\partial_{\mu}\pi_{1}^{3})\right]\\
                                        &\mathcal{L}^{(2)}_{\pi \pi V}=-\frac{\tilde{g}}{2}\epsilon_{abc}(\partial^{\mu}\pi_{1}^{a})W_{1\mu}^{b}\pi_{1}^{c}+\frac{g'}{2}\epsilon_{ab3}(\partial^{\mu}\pi_{1}^{a})\pi_{1}^{b}B_{2\mu}\\
                                        &\mathcal{L}^{(2)}_{\pi VV}=\frac{g'\tilde{g}F}{2}\epsilon_{ab3}W_{1}^{a\mu}B_{2 \mu}\pi_{1}^{b}\\
                                        &\mathcal{L}^{(2)}_{\pi \pi VV}=\mathcal{M}_{abcd}\tilde{g}^{2}\pi_{1}^{a}W_{1}^{b\mu}W_{1\mu}^{c}\pi_{1}^{d}-2\tilde{g}g'\mathcal{M}_{abc3}\pi_{1}^{a}W_{1}^{b\mu}\pi_{1}^{c}B_{2\mu}+\frac{g'^{2}}{8}B_{2\mu}^2(\pi_{1}^{a})^{2}\\
				\end{split}
			}
	\end{equation*}
where $M_{abcd}$ is given as before.

Turning to the kinetic energy term of $\Phi$, we see that its covariant derivative
\begin{equation}
D_{\mu}\Phi=\partial _{\mu}\Phi+igW_{0\mu}^1 \Phi-\frac{ig'}{2}B_{2\mu}\Phi
\end{equation}
can be expanded by plugging in Eq.~\eqref{eqn:phi representation}:

\begin{equation}
\Phi= \left( \begin{array}{c}
\frac{1}{\sqrt{2}}(\partial_{\mu}H+i\partial_{\mu}\pi_{t}^{0}) \\
i\partial_{\mu}\pi_{t}^{-} \end{array} \right)+ \frac{ig}{2}
\left( \begin{array}{c}
\frac{W_{0\mu}^{3}}{\sqrt{2}}(f+H+i\pi_{t}^{0})+\sqrt{2}iW_{0\mu}^{+}\pi_{t}^{-} \\
W_{0}^{-}(f+H+i\pi_{t}^{0})-iW_{0\mu}\pi_{t}^{-} \end{array} \right)-\frac{ig'}{\sqrt{2}}
\left( \begin{array}{c}
\frac{B_{2\mu}}{\sqrt{2}}(f+H+i\pi_{t}^{0}) \\
iB_{2\mu}\pi_{t}^{-} \end{array} \right).
\end{equation}

In order to make the expressions more compact, we will introduce the following notation:

\begin{align}
Z_{\mu}&=gW_{3\mu}-g'B_{2\mu}, \\ 
A_{\mu}&=gW_{3\mu}+g'B_{2\mu}.
\end{align}

The $Z$ and $A$ appearing in the above formulas are convenient aids to make the expressions look simple, and are \emph{not} the physical $Z_{\mu}$ and $A_{\mu}$. Using this, the product can be evaluated to be:

\begin{align}
D_{\mu}\Phi^{\dagger}D_{\mu}\Phi=&\frac{1}{2}(\partial_{\mu}H)^{2}+\frac{1}{2}(\partial_{\mu}\pi_{t}^{0})^{2}+(\partial^{\mu}\pi_{t}^{+})(\partial_{\mu}\pi_{t}^{-}) 
+\frac{Z_{\mu}}{2}\left[(f+H)(\partial_{\mu}\pi_{t}^{0})-\pi_{t}^{0}\partial_{\mu}H\right] \nonumber \\
&-\frac{g}{2}(\partial^{\mu}H) (W_{0\mu}^{+}\pi_{t}^{-}+W_{0\mu}^{-}\pi_{t}^{+})+\frac{ig}{2}(\partial^{\mu}\pi_{t}^{0})( W_{0\mu}^{+}\pi_{t}^{-}-W_{0\mu}^{-}\pi_{t}^{+}) +\frac{(Z_{\mu})^2}{8}\left[(f+H)^2+(\pi_{t}^0)^2\right] \nonumber \\
&+\frac{g}{2}\left[(f+H)[W_{0}^{-\mu}(\partial_{\mu}\pi_{t}^{+}) + W_{0}^{+\mu}(\partial_{\mu}\pi_{t}^{-})\right]+\frac{ig}{2}\pi_{t}^{0}\left[W_{0}^{-\mu}(\partial_{\mu}\pi_{t}^{+}) - W_{0}^{+\mu}(\partial_{\mu}\pi_{t}^{-})]\right] \nonumber \\
&+\frac{ig}{4}Z_{\mu}\left[(f+H)(W_{0}^{+\mu}\pi_{t}^{-}-W_{0}^{-\mu}\pi_{t}^{+})-i\pi_{t}^{0}(W_{0}^{+\mu}\pi_{t}^{-}+ W_{0}^{-\mu}\pi_{t}^{+})\right] \nonumber \\
&-\frac{iA_{\mu}}{2}\left[(\partial^{\mu}\pi_{t}^{+})\pi_{t}^{-}-(\partial^{\mu}\pi_{t}^{-})\pi_{t}^{+}\right]+\frac{g^2}{4}W^{-\mu}_{0}W_{0\mu}^{+}\left[(f+H)^2+(\pi_{t}^0)^2\right] \nonumber \\
&-\frac{ig}{4}A^{\mu}\left[(f+H)(W_{0\mu}^{+}\pi_{t}^{-}-W_{0\mu}^{-}\pi_{t}^{+})-i\pi_{t}^{0}(W_{0\mu}^{+}\pi_{t}^{-}+W_{0\mu}^{-}\pi_{t}^{+})\right] \nonumber \\
&+\frac{1}{2}\pi_{t}^{+}\pi_{t}^{-}(A_{\mu})^{2}+\frac{1}{2}\pi_{t}^{+}\pi_{t}^{-}W_{0}^{+\mu}W_{0\mu}^{-},
\label{eqn:phi lagrangian}
\end{align}
where $W^{\pm}=(W^{1}\mp iW^{2})/\sqrt{2}$, and similarly for the $\pi_{t}^{\pm}$. Eq.~\eqref{eqn:phi lagrangian} gives us the coupling of the top-Higgs and the pions to the gauge bosons, and the gauge-Goldstone mixing terms. Let us pick the latter contribution to the Lagrangian.

\begin{equation}
\mathcal{L}^{(3)}_{\textrm{mixing}} =\frac{gf}{2}\left[W_{0}^{-\mu}(\partial_{\mu}\pi_{t}^{+}) + W_{0}^{+\mu}(\partial_{\mu}\pi_{t}^{-}) \right]+\frac{f}{2}Z_{\mu}(\partial_{\mu}\pi_{t}^{0}).
\end{equation}
Plugging in the definitions of the fields, this becomes:

\begin{equation}
\mathcal{L}^{(3)}_{\textrm{mixing}}=\frac{f}{2}\left[g(\partial_{\mu}\pi_{t}^{a})W_{0}^{a\mu}-g'(\partial_{\mu}\pi_{t}^{3})B_{2}^{\mu}\right].
\end{equation}


\section{Four point couplings}

We present the four point couplings involving two gauge bosons and top-pions/top-Higgs in Table \ref{tab:H-couplings4}.

\begin{table}[!ht]
\begin{center}
\renewcommand{\arraystretch}{1.6}
  \begin{tabular}{cc }
   \hline \hline
Vertex & Strength \\
\hline
     $H_tH_tWW$ & $\frac{g_{0}^{2}}{4}\left(1+\frac{3\,x^2}{4}\right)$ \\ 
     $H_tH_tW'W$ & $-\frac{g_{0}^{2}x}{8}$ \\ 
     $H_tH_tW'W'$ & $\frac{g_{0}^{2}x^2}{16}$ \\ 
     $H_tH_tZZ$ & $\frac{e^{2}}{2}\left(\textrm{cosec}^{2}2\theta+\frac{x^2}{16}\left[1+2\,\textrm{cos}\,2\theta \right] \textrm{cosec}^{2}\theta\,\textrm{sec}^{4}\theta \right)$ \\ 
     $H_tH_tZ'Z$ & $-\frac{g_{0}^{2}x}{8}\textrm{sec}^{3}\theta\,\textrm{cos}\,2\theta$ \\ 
     $H_tH_tZ'Z'$ & $\frac{g_{0}^{2}x^2}{32}\textrm{sec}^{4}\theta\,\textrm{cos}^{2} 2\theta$ \\ 
     $H_tZW^{-}\Pi_{t}^{+}$ & $-\frac{i g_{0}^{2}}{2}\,\textrm{cos}\,\omega\,\textrm{tan}\,\theta\left(\textrm{sin}\,\theta+\frac{x^2}{16}\left[1+3\,\textrm{cos}\,2\theta \right]\textrm{sec}\,\theta\,\textrm{tan}\,\theta \right)$ \\ 
     $H_tZW^{'-}\Pi_{t}^{+}$ & $\frac{i g_{0}^{2}}{4}x\,\textrm{cos}\,\omega\,\textrm{sin}\,\theta\,\textrm{tan}\,\theta$ \\ 
     $H_tZ'W^{-}\Pi_{t}^{+}$ & $-\frac{i g_{0}^{2}}{4}x\,\textrm{cos}\,\omega \,\textrm{tan}^{2}\theta$ \\ 
     $H_tZ'W^{'-}\Pi_{t}^{+}$ & $\frac{i g_{0}^{2}}{8}x^2\,\textrm{cos}\,\omega \,\textrm{tan}^{2}\theta$ \\ 
     $H_tAW^{-}\Pi_{t}^{+}$ & $-\frac{i g_{0}^{2}}{2}\,\textrm{cos}\,\omega \,\textrm{sin}\,\theta\left(1+\frac{3\,x^2}{8} \right)$ \\
     $H_tAW^{'-}\Pi_{t}^{+}$ & $\frac{i g_{0}^{2}}{4}x\,\textrm{cos}\,\omega \,\textrm{sin}\,\theta$ \\ 
    \hline \hline
  \end{tabular}
  \caption{\label{tab:H-couplings4} Four point couplings involving the top-Higgs,
    again calculated to $\mathcal{O}(x^{2})$.}
\end{center}
\end{table}


\section{FCNC constraints and ideal delocalization}

\subsection{Limits on $\Delta \epsilon^2_{tL}$: $\Delta F=2$}

Limits on the deviation of
$\epsilon_{tL}$ from ideal comes from the minimal size of tree-level flavor-changing
neutral currents from $Z$-exchange. Consider re-writing Eq.~\eqref{eq:glzbb} for the
left-handed quarks of the $i$th family (where $i=u,c,t$) as
\begin{equation}
g_L^{Zii} = - \frac{e}{s_Wc_W} 
	\left[ \left(1 - \frac{\Delta \epsilon^2_{iL}}{2} \right) T_3
	- Q s_W^2 \right]~,
\end{equation}
where $\Delta \epsilon^2_{iL}$ denotes the deviation from ideal 
delocalization of the $i$th family in the top-quark mass-eigenstate basis
\begin{equation}
\Delta \epsilon^2_{iL} = \epsilon^2_{iL} - \frac{x^2}{2}~.
\end{equation}

In this notation  $t_L$ is the left-handed quark (in the top-quark mass eigenstate basis) whose ``down" component receives a large correction from top-pion exchange.  In general, this ``down" component may be written

\begin{equation}
d^t_{tL} = U_{3j} d_{jL}~,
\end{equation}
where $d^t_{iL}$ represent the ``down" components of the left-handed doublet
fields in the top-quark mass eigenstate basis, and $d_{jL}$ are the same fields
in the down-quark mass-eigenstate basis, and the $U_{3j}$ are the third row of
a unitary matrix. The minimal size of the $U_{3j}$, corresponding to ``next-to-minimal"
flavor violation~\cite{Agashe:2005hk}, is
\begin{equation}
U_{3j} = {\cal O}\left(V^{CKM}_{tj})\right),
\end{equation}
where $V^{CKM}$ is the usual CKM flavor-mixing matrix in the standard model.
Since GIM cancellation is exact when $\Delta \epsilon^2_{tL}=0$, we find
the tree-level flavor changing $Z$-boson couplings to down-quarks
\begin{eqnarray}
g^{Zb_L s_L}_Z = \frac{e \Delta \epsilon^2_{tL} V^{CKM}_{ts}}{4 s_W c_W}\label{eq:Zbs}\\
g^{Zb_L d_L}_Z = \frac{e \Delta \epsilon^2_{tL} V^{CKM}_{td}}{4 s_W c_W}\,.
\end{eqnarray}

\begin{figure}[th]
\centering
\includegraphics[width=0.60\textwidth]{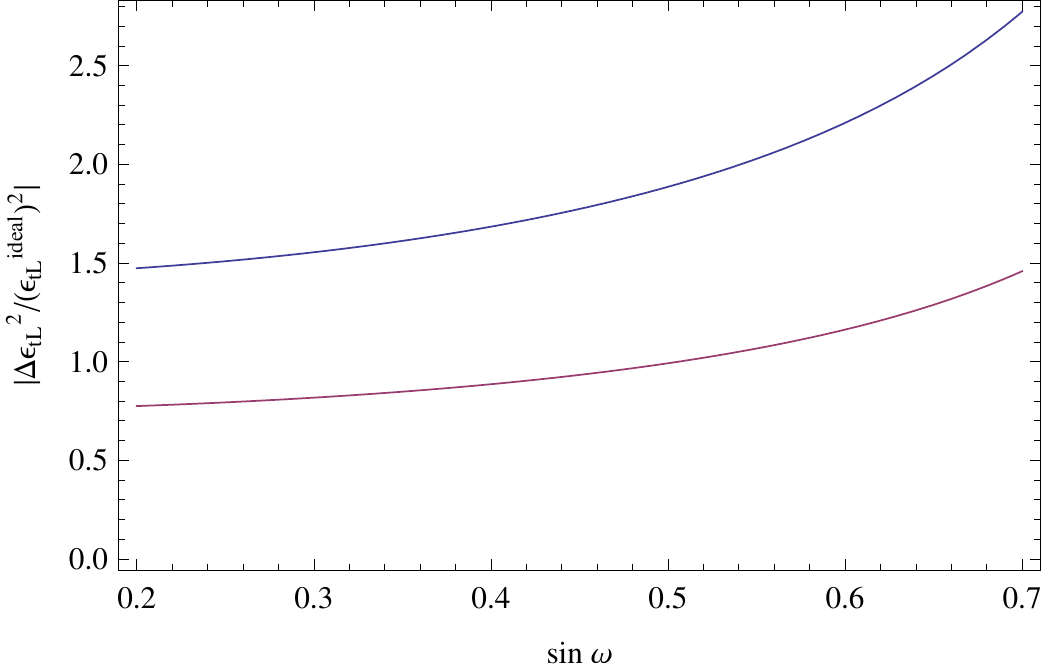}
\caption{The upper bounds in the deviation $|\Delta \epsilon^2_{tL}/(\epsilon^{ideal}_{tL})^2|$ arising
from limits on extra contributions to CP-violation in $K$-meson mixing (upper curve) and
from bounds on the rare decay $B_s \to \mu^+\mu^-$ (lower curve), as a function
of $\sin \omega$ for $M_{W'}=500$ GeV. 
\label{fig:two}
}
\end{figure}

$Z$-exchange then produces the $\Delta F=2$ effective operators
\begin{eqnarray}
C^1_K (\bar{s}_L \gamma^\mu d_L) (\bar{s}_L \gamma_\mu d_L) \label{eq:cdeffs}\\
C^1_{B_d} (\bar{b}_L \gamma^\mu d_L) (\bar{b}_L \gamma_\mu d_L) \\
\! C^1_{B_s} (\bar{b}_L \gamma^\mu s_L) (\bar{b}_L \gamma_\mu s_L) \,,\!
\end{eqnarray}
where, since GIM cancellation is exact when $\Delta \epsilon^2_{tL}=0$,  we find 
\begin{eqnarray}
|\Re (C^1_K)| = \left| \Re\left(\frac{e^2 (\Delta \epsilon^2_{tL}V^{CKM}_{ts} V^{CKM}_{td})^2}{8(s_W c_W)^2 M^2_Z}\right)\right| <\frac{1}{(1.0\times 10^6\, {\rm GeV})^2}\\
|\Im (C^1_K)| = \left| \Im\left(\frac{e^2 (\Delta \epsilon^2_{tL}V^{CKM}_{ts} V^{CKM}_{td})^2}{8(s_W c_W)^2 M^2_Z}\right) \right| <\frac{1}{(1.5\times 10^7\, {\rm GeV})^2}\\
|C^1_{B_d}| =\left|  \frac{e^2 (\Delta \epsilon^2_{tL}V^{CKM}_{tb} V^{CKM}_{td})^2}{8(s_W c_W)^2 M^2_Z}\right| <  \frac{1}{(2.1\times 10^5\, {\rm GeV})^2}\\
|C^1_{B_s}| =\left|  \frac{e^2 (\Delta \epsilon^2_{tL}V^{CKM}_{tb} V^{CKM}_{td})^2}{8(s_W c_W)^2 M^2_Z}\right|< \frac{1}{(3 \times 10^4\, {\rm GeV})^2} \,,\!
\end{eqnarray}
where the bounds given by the last inequality in each expression come from Ref. \cite{Bona:2007vi}.
The strongest constraint arises from limits on extra contributions to CP-violation in K-meson
mixing, for which we find
\begin{equation}
\Delta \epsilon^2_{tL} < 7.2 \times 10^{-2}~.
\label{eq:CPlimit}
\end{equation}
We plot this bound as a limit on $\Delta \epsilon^2_{tL}/(\epsilon^{ideal}_{tL})^2$ in the
upper curve in Fig.~\ref{fig:two}, 
as a function of $\sin\omega$ for $M_{W'}=500$ GeV. 

\subsection{Limits on $\Delta \epsilon^2_{tL}$: $\Delta F=1$}

The strongest limits from $\Delta F=1$ processes come from constraints
on the $B$-meson decays $B_{d,s} \to \mu^+\mu^-$. The limits arising from
experimental constraints have been summarized in \cite{Carpentier:2010ue}.
In Table 2 of that reference, we find the strongest constraint arising from 
Tevatron limits on $B_s \to \mu^+ \mu^-$ and, for the operator
\begin{equation}
\frac{2\varepsilon}{v^2} (\bar{b}_L \gamma^\nu s_L)(\bar{\mu}_L \gamma_\nu \mu_L)~,
\label{eq:varepsilon}
\end{equation}
where $\sqrt{2} G_F = v^{-2}$ and $v \approx 246$ GeV is the weak scale.
In our case, using Eq.~\eqref{eq:Zbs} and $M_Z = e v/(2 s_W c_W)$, we find
\begin{equation}
\varepsilon = \frac{1}{2}\Delta \epsilon^2_{tL} V^{CKM}_{ts} V^{CKM}_{tb} \approx \frac{\lambda^2 \Delta\epsilon^2_{tL}}{2}~.
\end{equation}

Using the limit $BR(B_s \to \mu^+ \mu^-)< 4.3 \times 10^{-8}$ \cite{CDFnew} and the techniques\footnote{We
disagree with the numerical extraction of the bound on $\varepsilon$ presented in \protect\cite{Carpentier:2010ue},
though we agree with their method.} of ref. \cite{Carpentier:2010ue}, we find the bound
$\varepsilon < 7.6 \times 10^{-4}$. From eqn. (\ref{eq:varepsilon}), we then obtain
\begin{equation}
\Delta \epsilon^2_{tL} < 3.8 \times 10^{-2}~,
\end{equation}
a constraint roughly twice as small as that given by limiting contributions to
CP-violation in $K$-meson mixing in Eq.~\eqref{eq:CPlimit}.

Comparing Figs.~\ref{fig:one} and \ref{fig:two}, we see that compensating for the deviation in $R_b$ resulting
from top-pion exchange by modifying the delocalization of the third-generation quarks is not, in the context of ``next-to-minimal" flavor violation~\cite{Agashe:2005hk}, ruled out by flavor changing neutral current constraints.


\end{document}